\documentclass[preprint2]{aastex}  

\shorttitle{SIGNATURES OF LONG-LIVED SPIRAL PATTERNS}
\shortauthors{Mart\'{\i}nez-Garc\'ia et al.}

\usepackage{amsmath} 

\begin{document}

\title{SIGNATURES OF LONG-LIVED SPIRAL PATTERNS.}

\author{Eric E. Mart\'inez-Garc\'ia\altaffilmark{1,2}}
\affil{1 Instituto Nacional de Astrof\'isica, \'Optica y Electr\'onica (INAOE), Aptdo. Postal 51 y 216, 72000 Puebla, Pue., M\'exico.\\
2 Instituto de Astronom\'ia, Universidad Nacional Aut\'onoma de M\'exico, AP 70-264, Distrito Federal 04510, M\'exico.}
\email{ericmartinez@inaoep.mx; martinez@astro.unam.mx}

\and

\author{Rosa A. Gonz\'alez-L\'opezlira\altaffilmark{3,4,5,6}}
\affil{3 Centro de Radioastronom\'ia y Astrof\'isica, UNAM, Campus Morelia,
     Michoac\'an, M\'exico, C.P. 58089}
\email{r.gonzalez@crya.unam.mx}

\altaffiltext{4}{Visiting astronomer at Kitt Peak National Observatory, National Optical
Astronomy Observatory, which is operated by the Association of Universities for
Research in Astronomy (AURA), under cooperative agreement with the National
Science Foundation.}
\altaffiltext{5}{Visiting astronomer at Cerro Tololo Inter-American
Observatory, National
Optical Astronomy Observatory, which is operated by the AURA, under contract
with the National Science Foundation.}
\altaffiltext{6}{Visiting astronomer at Lick Observatory, which is
operated by the
University of California.}

\begin{abstract}

Azimuthal age/color gradients across spiral arms are a signature of long-lived spirals.
From a sample of 19 normal (or weakly barred) spirals
where we have previously found azimuthal age/color gradient candidates, 13 objects
were further selected if a two-armed grand-design pattern survived in a 
surface density stellar mass map. Mass maps were obtained from optical and
near-infrared imaging,  
by comparing with a Monte Carlo library of stellar population synthesis models
that allowed us to obtain the mass-to-light ratio in the $J$-band, $(M/L)_J$, as a function of
($g-i$) versus\ ($i-J$) color.
The selected spirals were analyzed with Fourier methods in search for other signatures of long-lived modes
related to the gradients, such as the gradient divergence toward corotation, and the behavior
of the phase angle of the two-armed spiral in different wavebands, as expected from theory.
The results show additional signatures of long-lived spirals in at least
50\% of the objects. 

\end{abstract}


\section{Introduction.}

Understanding the origin of spiral arms, bars, and rings is key for discerning the long-term 
evolution of disk galaxies. The density wave (DW) theory, originally conceived by~\citet{lind63} and developed
in a linear form by~\citet{lin64}, has been commonly brought up to explain spiral structure.
One fundamental prediction of this theory is the presence of offsets between
tracers of the different stages of star formation and
evolution~\citep{rob69}.
A manifestation of these offsets would be age/color gradients
across arms, as a
result of the interaction between a spiral pattern
with an angular speed that is nearly constant with radius, and stars and
gas with differential rotation. 
The gas overtakes the spiral pattern inside corotation
(CR, i.e., the radius where the pattern and orbital angular
velocities are the same), and is left behind outside CR.
As gas feels the gravitational potential near the spiral arms, it shocks~\citep{rob69,git04}.
This produces the piling up of dust and molecular gas that is commonly found in the concave side of the
arms, if the process occurs inside CR and spiral arms trail.
If the shock triggers star formation, an age (or color) gradient is expected across spiral
arms as the new stars age and drift away from the arms. Young stars would be located immediately 
downstream the dust lanes, followed by increasingly older stars in the direction
of rotation. In part due to the difficulties
encountered to detect such gradients, the relevance of spiral DWs
for disk galaxies has been repeatedly challenged. Recent examples are
\citet{foy11},~\citet{gra12a,gra12b}, and~\citet{fer12}.

In a series of papers, we have detected and studied azimuthal color/age gradients across spiral arms and bars, thus 
establishing a relation between spiral or bar dynamics and star formation~\citep{gon96,mar09a,mar09b,mar11}.
Our studies suggest that azimuthal gradients 
can be found in almost every galaxy, but are difficult to detect and 
do not appear widespread within a single galaxy. 
\citet{efr10} has noticed an inverse correlation between the signatures of a spiral shock,
and the presence of chains of star complexes (``beads on a string'') in spiral arms;
he has also observed that spiral shocks seem to be associated with  
an irregular magnetic field~\citep{dobpr08},
whereas the ``beads on a string" configuration correlates with
the existence of a regular magnetic field.\footnote{
 On smaller scales than color gradients,~{\it{Herschel}} Space Observatory~\citep{pil10} has shown that
stars form along gaseous filaments, whose width
is surprisingly constant, $\sim 0.1$ pc, irrespective of central column density~\citep{arz11}.}

The ``infant mortality'' of star clusters~\citep{lad03,goo06},
that leads to their disruption in the first 10-30 Myr,
may prevent the formation of age gradients even if shocks trigger star formation.
Stellar winds and supernovae may drive out the gas not used in star formation,
causing a change in the gravitational potential. This would lead to the disintegration
of the clusters, as stars become unbound.
Coherent age gradients must first survive the ``infant mortality'' process;
then, after $t_{\rm age} > 50$ Myr, the ``dissolution of stellar groups''
scenario proposed by~\citet{wie77} can also take place near spiral arms.
These last process can explain the apparent ``downstream decline'' of
the gradients already noticed by us~\citep[cf.][]{mar09a,mar11}.

\subsection{Pattern speed variation with Radius.}~\label{OmegaVar}

Recently, based on the analysis with the radial Tremaine-Weinberg (RTW)
method of CO and HI data of several galaxies,
it has been proposed that the spiral pattern speed $\Omega_{\rm p}$ may
actually increase with decreasing radius in some objects~\citep{mer06,mei09,spe12}.
According to~\citet{wada11} and~\citet{gra12a,gra12b}, this behavior
is also seen in simulations.\footnote{\citet{roc13} find
that a radially varying $\Omega_{\rm p}$ occurs only in simulations with
flocculent, transient arms.} 
If this is indeed the case, there would be no real long-lived
pattern with solid body rotation, and hence 
no significant offsets between the tracers of 
the different stages of star formation would be expected
\citep[see, for example,][]{gra12b}. In particular,
azimuthal color gradients across the arms would not be observed.

\begin{figure*}
\centering
\epsscale{2.0}
\plotone{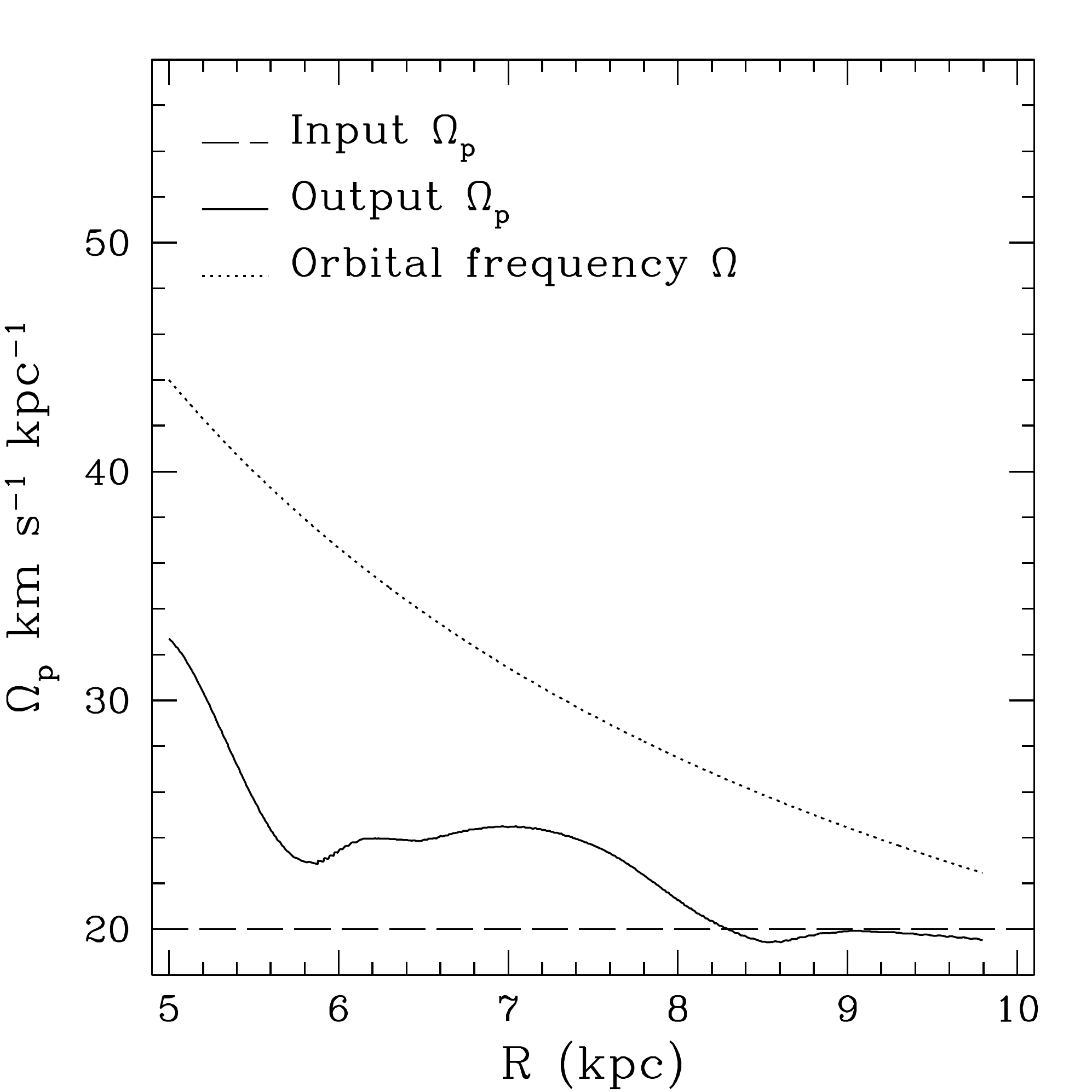}
\caption[f1]{Pattern speed, $\Omega_{\mathrm{p}}$ (km s$^{-1}$ kpc$^{-1}$), vs. radius, $R$ (kpc), obtained from the MHD simulation
analyzed in~\citet{mar09b}. Long-dashed line: input $\Omega_{\mathrm{p}}$; solid line: pattern
speed values obtained by applying the GG96 method to synthetic azimuthal color(age) gradients across the arms,
under the circular motion assumption; dotted line: orbital frequency $\Omega$.
~\label{nonCIRCULAR}}
\end{figure*}

In~\citet{mar09b}, we investigated the effect of having (and neglecting) non-circular streaming motions, 
that is, of using  
a purely circular dynamic model to derive 
spiral pattern speeds from observations of color gradients.
Summarizing, based on semi-analytical solutions and MHD simulations, gas orbits were obtained
for models with a fixed pattern speed. Stellar population synthesis (SPS) models of
age gradients were incorporated to these gas orbits, and synthetic
photometric observations were performed.\footnote{We assumed that
young stars retain the velocity
components of their parental molecular clouds.}
The method of~\citet[][hereafter GG96]{gon96} was then applied to measure $\Omega_{\rm p}$. 
In spite of having a real pattern speed, constant with radius, 
higher pattern speeds were measured at smaller radii (see Figure~\ref{nonCIRCULAR}).
As mentioned in~\citet{mar09b},
the reason for this bias
is that gas streamlines in a steady rotating spiral shock
turn somewhat along the arm after passing through the shock.
Hence, stars take longer to move away from the arm, 
and the observer will think there is a smaller difference between 
the pattern speed and the orbital frequency. $\Omega_{\rm p}$
will seem to follow $\Omega$, just like in the previously
mentioned applications of the radial RTW method, except that
in our case we can be sure that the trend is caused by 
the streaming motions of the young stars, rather than by a radial
variation of $\Omega_{\rm p}$. 
The bias is also found in real observations, as shown in~\citet{mar09b}.

In an ideal scenario where $\Omega_{\rm p}$ is indeed constant for all radii and 
spiral arms trigger star formation
in all their extension, color gradients are predicted to be located
all along the arms (except near CR). 
In reality, age gradients are detected only in a few small regions
per galaxy, but they do seem linked to disk dynamics
\citep[GG96;][]{mar09a,mar09b,mar11}. So far, we have detected gradients through visual inspection. 
In order to locate gradients in a more objective way, and to 
attempt to identify some that do not meet the theoretical
expectations of coherence or smoothness, 
in what follows we will adopt Fourier techniques to study normal (or weakly barred) spirals.

\section{Photometric data.}

Our data consist of deep photometric images of the 13 SA and SAB spirals\footnote{
NGC~578, NGC~918, NGC~1417, NGC~1421, NGC~3162, NGC~3938, NGC~4254,
NGC~4939, NGC~5371, NGC~6951, NGC~7125, NGC~7126, and NGC~7753.
As argued in~\citet{mar09a}, NGC~578 is peculiar, in the sense that it is the only object in the
sample whose arms may end at CR and not at the outer Lindblad 
resonance (OLR). We also caution the reader that NGC~1421 is seen nearly edge-on,
with an inclination angle $\alpha \sim 76 \degr$.}
described in~\citet{mar09a}.
The data include mosaics in the optical $g$, $r$, and $i$ bands~\citep[][see Table~\ref{tbl-filters}]{thu76,wade79},
the near-infrared (NIR) $J$ band, and the NIR $H$, $K$, $K_s$ or $K^\prime$ filters.
We have now added six new objects,\footnote{NGC~1703 (type SB), NGC~3001 (SAB), NGC~3338 (SA), 
NGC~4603 (SA), NGC~6907 (SB), and NGC~7083 (SA). All galaxy types are taken from the Third Reference 
Catalogue of Bright Galaxies~\citep[RC3,][]{dev91}.} whose observation log is shown in Table~\ref{tbl-obslog}.
NGC~1703 and NGC~6907, classified as SB by RC3, were included in the analysis because (apparently) 
they have no prominent bar structure, as appreciated in the images.
The data were reduced and calibrated with the same standard procedures used in~\citet{mar09a},
and~\citet{mar11}.
The optical ($g$, $r$, and $i$) calibration was done in the Thuan-Gunn system~\citep{thu76,wade79}.
The zero point of this photometric system is chosen such that
the standard star BD+17$\degr$4708 has $g=r=i=9.5$ mag.
The NIR $J$ data were calibrated with images from the Two Micron All Sky Survey~\citep[2MASS;][]{skr06}.


\begin{deluxetable}{ccc}
\tabletypesize{\scriptsize}
\tablecaption{Filter Characteristics~\label{tbl-filters}}
\tablewidth{0pt}
\tablehead{
\colhead{Filter} & \colhead{$\lambda_{\rm eff}$} & \colhead{FWHM}
}
\startdata
\emph{$g$}       & 5000\AA\AA & 830\AA\AA \\
\emph{$r$}       & 6800\AA\AA & 1330\AA\AA \\
\emph{$i$}       & 7800\AA\AA & 1420\AA\AA \\
\emph{$J$}       & 1.25\micron & 0.29\micron \\
\enddata

\tablecomments{
Columns 2 and 3: effective wavelengths and widths, respectively, of the filters 
mainly used for analysis.
}

\end{deluxetable}


\begin{deluxetable}{cccccl}
\tabletypesize{\scriptsize}
\tablecaption{Observation Log and Galaxy Parameters\label{tbl-obslog}}
\tablewidth{0pt}
\tablehead{
\colhead{Object}
& \colhead{Filter}
& \colhead{Exposure (s)}
& \colhead{Telescope}
& \colhead{Date (month/year)}
& \colhead{* Parameters}
}

\startdata

NGC~1703            & $g$     & 5100 & CTIO 0.9 m       & 3/94, 3/95   &  Type: SB(r)b    \\
                    & $r$     & 4200 &    ''            &  ''          &  PA (deg): 104\tablenotemark{a}   \\
                    & $i$     & 5400 &    ''            &  ''          &  $\alpha$ (deg): 27.0 $\pm$ 12.1\\
                    & $J$     & 2717 & CTIO 1.5 m       & 2/94         &  $v_{\rm{rot}}$ (km s$^{-1}$): 55 $\pm$ 20\\
                    & $K$     & 1281 &    ''            &  ''          &  Dist (Mpc):  20.4 $\pm$ 1.7 \\
                    & $H$     & 2758 &    ''            &  ''          &  \\
\tableline

NGC~3001            & $g$     & 4500 & CTIO 0.9 m       & 3/94, 3/95   &  Type: SAB(rs)bc  \\
                    & $r$     & 5100 &    ''            &  ''          &   PA (deg): 6     \\
                    & $i$     & 4500 &    ''            &  ''          &   $\alpha$ (deg): 47.5 $\pm$ 4.9\\
                    & $J$     & 2718 & CTIO 1.5 m       & 2/94         &   $v_{\rm{rot}}$ (km s$^{-1}$): 248 $\pm$ 20\\
                    & $H$     & 2700 &    ''            &  ''          &   Dist (Mpc): 35.6 $\pm$ 3.0 \\

\tableline

NGC~3338            & $g$     & 3600 & Lick 1 m         & 4/94, 11/94         &   Type: SA(s)c  \\
                    & $r$     & 4500 &    ''            &  ''                 &   PA (deg): 100 \\
                    & $i$     & 4200 &    ''            & 2/94, 4/94, 11/94   &   $\alpha$ (deg): 51.9 $\pm$ 2.1\\
                    & $J$     & 1200 & Lick 1 m         & 12/94, 2/95         &   $v_{\rm{rot}}$ (km s$^{-1}$): 194 $\pm$ 9\\
                    & $J$     & 1002 & Kitt Peak 1.3 m  & 3/94, 11/94         &   Dist (Mpc): 23.7 $\pm$ 2.1 \\
                    & $K_{s}$ & 926  &    ''            &  ''                 &  \\

\tableline

NGC~4603            & $g$     & 5400 & CTIO 0.9 m       & 3/94, 3/95          &   Type: SA(s)c  \\
                    & $r$     & 4500 &    ''            &  ''                 &   PA (deg): 27 \\
                    & $i$     & 4200 &    ''            &  ''                 &   $\alpha$ (deg): 43.6 $\pm$ 5.7 \\
                    & $J$     & 1393 & CTIO 1.5 m       & 2/94                &   $v_{\rm{rot}}$ (km s$^{-1}$): 242 $\pm$ 32\\
                    & $H$     & 1399 &    ''            &  ''                 &   Dist (Mpc): 29.4 $\pm$ 2.6 \\

\tableline

NGC~6907            & $g$     & 3600 & CTIO 0.9 m       & 9/94         &   Type: SB(s)bc   \\
                    & $r$     & 3900 &    ''            &  ''          &   PA (deg): 46 \\
                    & $i$     & 3600 &    ''            &  ''          &   $\alpha$ (deg): 35.6 $\pm$ 3.7 \\
                    & $J$     & 570  & CTIO 1.5 m       & 9/94, 9/95   &   $v_{\rm{rot}}$ (km s$^{-1}$): 244 $\pm$ 41\\
                    & $K_{s}$ & 355  &    ''            &  ''          &   Dist (Mpc): 47.7 $\pm$ 4.1 \\

\tableline

NGC~7083            & $g$     & 3600 & CTIO 0.9 m       & 9/94                &   Type: SA(s)bc  \\
                    & $r$     & 3900 &    ''            &  ''                 &   PA (deg): 5 \\
                    & $i$     & 3900 &    ''            &  ''                 &   $\alpha$ (deg): 52.9 $\pm$ 3.0 \\
                    & $J$     &  840 & CTIO 1.5 m       & 9/94                &   $v_{\rm{rot}}$ (km s$^{-1}$): 226 $\pm$ 10\\
                    & $K_{s}$ &  392 &    ''            &  ''                 &   Dist (Mpc): 45.4 $\pm$ 3.9 \\
                                                                                                 
\enddata

\tablecomments{
Column 6-galaxy parameters: Hubble types from the RC3; 
P.A.: position angles from RC3;
inclination angle, $\alpha = {\rm cos}^{-1}{(b/a)}$, where $a/b$ is the isophotal diameter ratio derived from the $R_{25}$ parameter in RC3;
$v_{\rm{rot}}$: galactic rotation velocity obtained from the HI data of~\citet{pat03}, corrected for inclination;
dist: Hubble distance obtained from the RC3 heliocentric radial velocity and the infall model of~\citet{mou00}, $H_0=71\pm6$ km s$^{-1}$ Mpc$^{-1}$.
}

\tablenotetext{a}{~\citet{pat00}.}

\end{deluxetable}


\section{Structural type of the spiral arms}

The DW theory predicts the existence of age gradients across spiral arms.
Multi-armed or flocculent galaxies, however, are not described by this theory.
The short arms of the latter are probably unbound star complexes sheared by galactic differential rotation, 
whereas the ``knee-like''\footnote{Consisting of straight fragments that join at angles of roughly $135\degr$.}
shaped arms of M101-type galaxies might be formed by the gravitational instability of galactic disks;
if so, they are transient features~\citep{cla06,dobbo08}.
The azimuthal profiles (intensity versus $\theta$) of such arms should be symmetric about the gravitational potential well, 
where the gas is yet to be transformed into stars; this configuration is indeed sometimes observed in
certain galaxies (for example, in NGC~4535 and M61, 
a narrow dark lane is seen right at the center of a stellar arm). 

To analyze the validity of the DW theory, one should 
study galaxies whose structure is likely to be explained by it. 
Mainly, from the point of view of their 
spiral arms, there are three different types of galaxies~\citep[e.g.,][]{efr11}.
These are (1) symmetric grand-design spirals (e.g., M81), with mass arms 
due to DWs; (2) multi-armed or 
``knee-like'' spirals (e.g., M101);
and (3) flocculent spiral galaxies (e.g., NGC~2841). However, even in the same galaxy, 
different arms may be formed by different mechanisms.
Thus, in the classical multi-armed galaxy 
M101 the inner arms 
are rather regular, and in our own Galaxy the quite symmetrical inner arms transform into 
a polygonal multi-armed structure in the outer regions~\citep{efr11}.
Albeit flocculent and multi-armed structures have a low mass-to-light ratio
($M/L$) and are seen mainly in the optical
(see, for example, the classical example of NGC~309 in Block et al.\ 1994), 
in order to determine whether the arms in the NIR are indeed mass DWs
it is important to disentangle the 
contribution of young stars and clusters at these longer wavelengths;
such contribution can reach up to 20\%-30\%
(e.g., Rix \& Rieke 1993; GG96; Rhoads 1998; James \& Seigar 1999; 
Patsis et al.\ 2001; Grosb{\o}l et al.\ 2006).
With the aim of determining the actual structural type of the spiral arms, 
resolved stellar mass maps are required. Mass maps get around the problem that the
$M/L$, at any wavelength, is not constant across the disk and near spiral arms.

\begin{figure*}
\centering
\epsscale{2.0}
\plotone{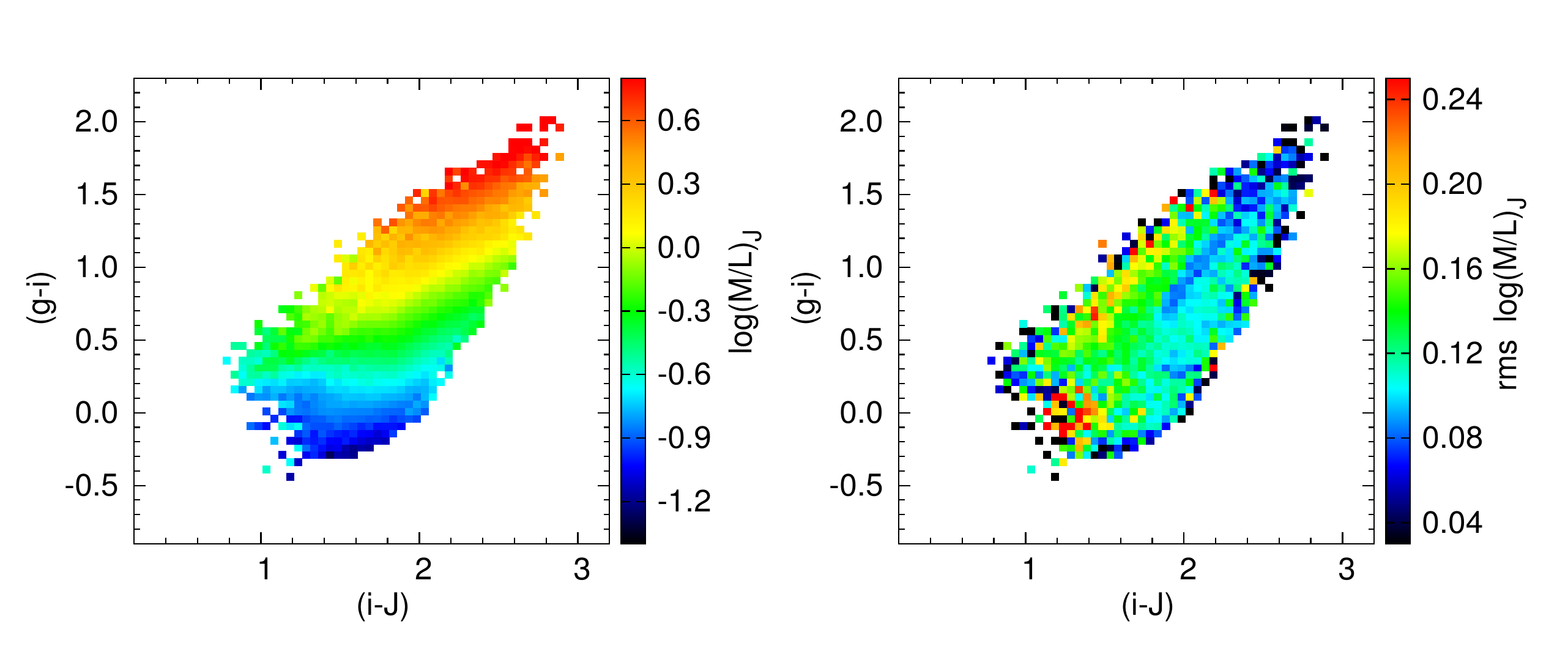}
\caption[f2]{Mass-to-light ratios ($M/L$s) as a function of the color-color space ($g-i$) vs. ($i-J$).
Left panel: median logarithmic $M/L$ in the $J$ band, $(M/L)_J$, for models binned 
in elements with size $0.05 \times 0.05$ mag$^2$ in two-dimensional color.
Right panel: rms of log $M/L$.
$M/L$ is the ratio between stellar mass and the light that reaches the observer,
i.e., the~{\it{effective}} $M/L$~\citep{zib09}. 
Optical $g$ and $i$ photometry in the Thuan-Gunn system~\citep{thu76,wade79}.
$J$ band calibrated as in the 2MASS survey~\citep{skr06}.}~\label{colorsML}
\end{figure*}

In order to build resolved maps of stellar mass of our sample galaxies, we use 
the method of~\citet{zib09}. 
The method relies on a Monte Carlo library of 50,000 stellar population spectra, constructed from the SPS models 
of~\citet{bc03}, and Charlot~\& Bruzual (2007, private communication). Each spectrum
is computed by randomly drawing the model parameters (star formation history, metallicity,
and dust attenuation\footnote{Treated as prescribed in the two-component
dust model of~\citet{cha00}.})
from adequate physical distributions~\citep[see also][]{daC08}.
We use the ``MAGPHYS'' code\footnote{http://www.iap.fr/magphys/magphys/MAGPHYS.html.}
of~\citet{daC08} to extract colors from the spectral library.
Each element in a theoretical two-dimensional color-color plot
can be produced by several combinations of model parameters.   
The median $M/L$ of all these combinations is then found, and
attributed to the data with the same position in the observed two-dimensional color-color plot.
Resolution elements in two-dimensional color space (in our case, $g-i$ versus $i-J$)
have a size of $0.05\times 0.05$ mag$^2$,
in order to account for the observational error. We compute the median $M/L$
in the $J$-band (see Figure~\ref{colorsML}), and compare the resulting distributions
with the colors obtained from the observations,\footnote{
All images are treated with the adaptive smoothing code ``ADAPTSMOOTH'', developed by Stefano Zibetti~\citep{zib09b},
and available at the URL http://www.arcetri.astro.it/$\sim$zibetti/Software/
ADAPTSMOOTH.html
The code enhances the S/N with a minimum loss of effective resolution, and keeps the photometric fluxes unaltered.
The models are reddened before comparing with the observations, 
to take into account
foreground Galactic extinction~\citep{sch98}.} on a pixel-by-pixel basis.
This comparison produces a resolved map of $(M/L)_J$. Finally, a relative ``mass map" is obtained, 
equal to the product of the $(M/L)_J$ map by the  
$J$-band image. Absolute mass maps can be worked out, taking into account the distance to the galaxies,
but are not necessary for our purposes.

\begin{figure*}
\centering
\epsscale{2.0}
\plotone{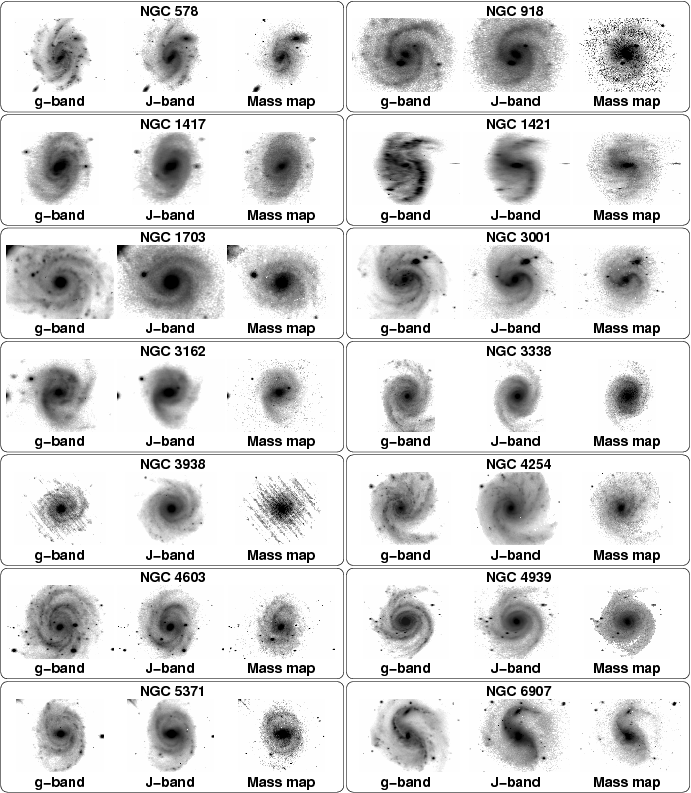}
\caption[f3]{Deprojected $g$ band, NIR $J$ band, and resolved maps of stellar mass (``mass maps'',
obtained via the technique of Zibetti et al.\ 2009). The images are displayed in logarithmic scale. 
Foreground stars and other objects were not removed in order to preserve spatial reference points.
~\label{J_vs_Mass}}
\end{figure*}

\begin{figure*}
\centering
\epsscale{2.0}
\plotone{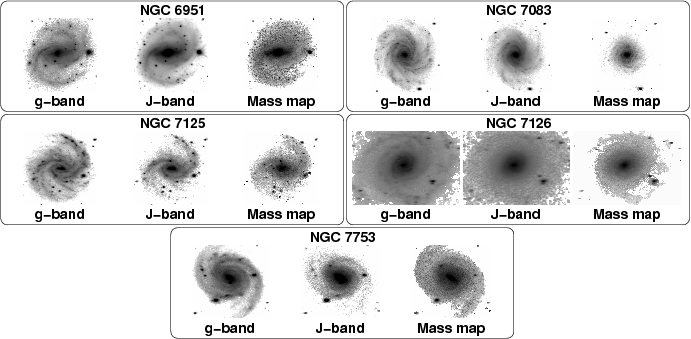}
\caption[f4]{Figure~\ref{J_vs_Mass}, continued.
~\label{J_vs_Mass_cont}}
\end{figure*}

The results are shown in Figures~\ref{J_vs_Mass} and~\ref{J_vs_Mass_cont}, where
we compare the $g$ and $J$ frames with the mass maps at the same relative scale.
We then proceed to select the objects in our sample that retain a spiral
structure that may be explained by the DW theory. From this {\it{visual}}\footnote{
Future works will include a more objective selection procedure.}
inspection we reject the following objects prior to the Fourier analysis:

\begin{enumerate}

\item~{\it{NGC~3162}}. There is no two-armed spiral structure appreciable in the mass map.

\item{\it{NGC~3938}}. The mass map has a very low quality, due to
a low signal-to-noise ratio (S/N) of the $g$-band mosaic.

\item{\it{NGC~5371}}. Ring structures rather than spirals are discerned in the mass map.

\item{\it{NGC~7083}}. The spiral structure looks flocculent in the mass map.

\item{\it{NGC~7126}}. No clear spirals can be appreciated in the mass map.

\item{\it{NGC~7753}}. Spiral structure is not observed in the mass map.

\end{enumerate}

The remaining 13 objects were analyzed as described below.

\section{Fourier analysis of the gradients.}~\label{Fourier}

\subsection{Azimuthal $Q(rJgi)$ Index Analysis.}~\label{Q_fourier}

The first step in our Fourier analysis consists in generating $Q(rJgi)$ mosaics,
following the method of GG96.
$Q$ is defined as follows:

\begin{equation}
  Q(rJgi) = (r-J) - \frac{E(r-J)}{E(g-i)}(g-i),
\end{equation}

\noindent where $\frac{E(r-J)}{E(g-i)}$ is the color excess term.
This photometric index is reddening-free for screen absorption, and also for a
mixture of dust and stars, as long as $\tau_{V} < 2 $.
At the same time, $Q(rJgi)$ is sensitive to supergiant stars, and therefore traces star formation.
SPS models~\citep{bc03} predict that, 
following a continuous burst of star formation that lasts for $\sim 2 \times 10^7$ yr,
the $Q(rJgi)$  index increases its value during $\sim 2.6 \times 10^7$ yr
and then starts to decline~\citep[cf., GG96;][]{mar09a}.\footnote{These models were computed for a
mixture of young ($\sim 2 \%$ by mass) and old stars
(with an age of $5 \times 10^9$ yr). Both populations have solar metallicity and a Salpeter 
initial mass function (IMF), with $M_{\rm lower} =$ 0.1 $M_\odot$, 
and $M_{\rm upper} =$ 10 $M_\odot$. The choice of $M_{\rm upper}$ is justified by the
observed inverse correlation between the locations where color gradients are detected,
and those with important $H_\alpha$ emission~\citep[GG96;][]{mar09a,mar11}.}
$Q(rJgi)$ mosaics of our sample were transformed to polar coordinates.\footnote{$\theta$ vs.\ ln $R$; logarithmic
arms appear as straight lines in such a map, with slope $m = {\rm cot}\ (-i)$, where $i$ is the 
arm pitch angle.} Spiral arms
were then ``straightened'', by adding to the $\theta$ coordinate a phase
as a function of $\ln R$,  until the arms appear horizontal.
Unlike our previous studies, where the ``straightening slope'' was different
depending on the specific arm and region under
analysis, here we adopt the same ``straightening slope'' for both (or all the) arms in the same
object. For each galaxy, this slope is obtained from the median of the  pitch angles 
measured in different
wavelengths~(see Section~\ref{pitch_angles} and Table~\ref{tbl-pitchs}).\footnote{
We define the median pitch angle for each object by arranging the angles in 
$g$, $r$, $i$, and $J$ 
from the lowest to the highest value, and averaging together the two middle values.}

\begin{figure*}
\centering
\epsscale{2.0}
\plotone{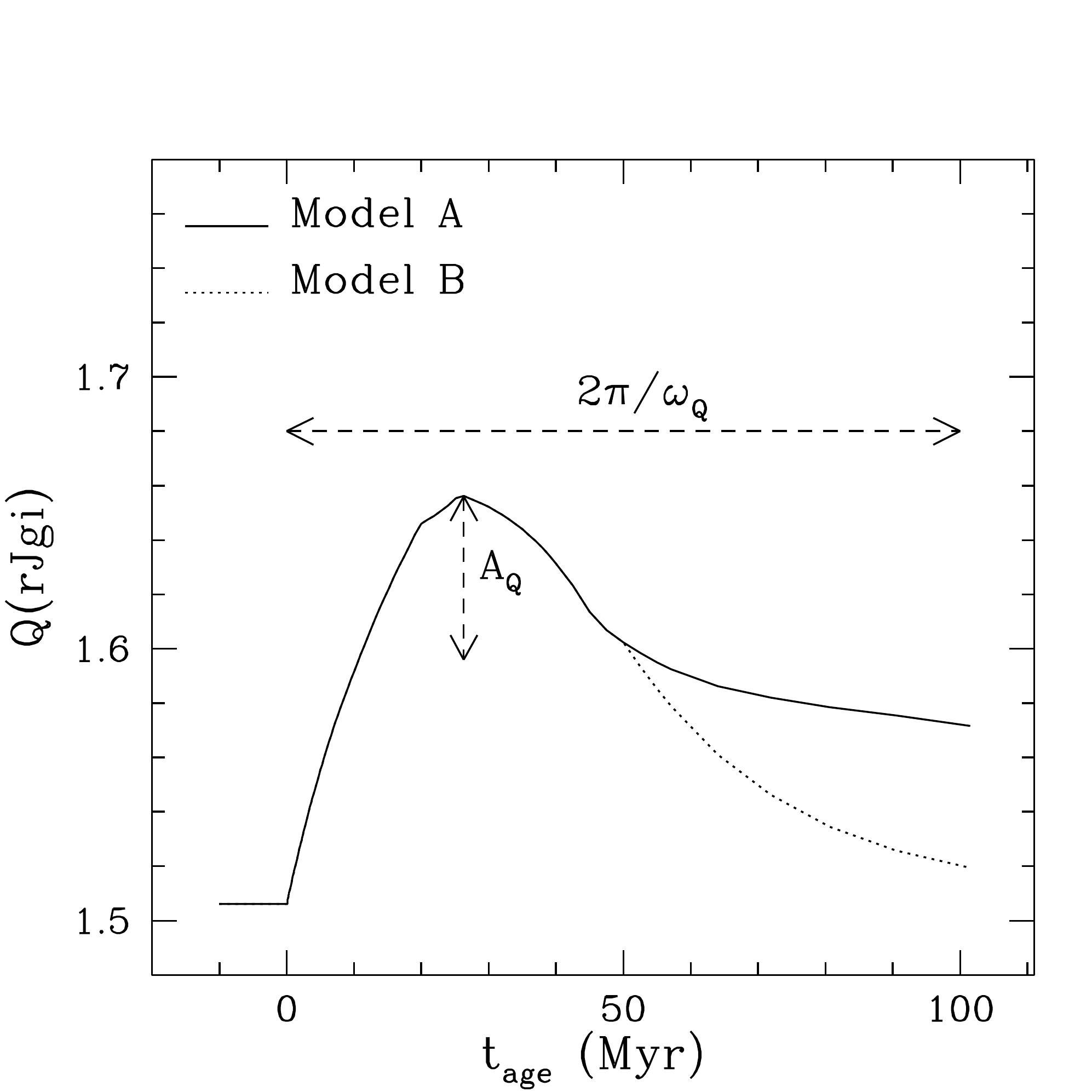}
\caption[f5]{$Q(rJgi)$ index vs. stellar population age for models ``A'' (solid line),
derived from SPS models only, and ``B'' (dotted line),
including the ``dissolution of stellar groups'' after 50 Myr~\citep{wie77,mar11}. 
The $Q$ profile can be characterized, using Fourier methods,
as a nearly sinusoidal signal with angular frequency $\omega_Q$, and amplitude $A_{Q}$
(e.g., compare model ``B'' with a sine function, $\sin{(\theta)}$, valuated from $-\pi/2$ to $3\pi/2$;
in such case, $\omega_Q=1$).
~\label{Qmodel}}
\end{figure*}

We divide the ``straightened'' images in vertical sections, equivalent to concentric annuli 
in a deprojected image of the galaxy, in order to perform a Fourier analysis.\footnote{
We ran some tests that demonstrate that the results are
independent of the width of the annuli. Widths were chosen to be the 
same as in~\citet{mar09a}.} 
Here we aim to compare the observed $Q(rJgi)$ index profiles with the
SPS models adopting Fourier techniques.
SPS models give $Q(rJgi)$ as a function of age, $t_{\rm age}$, while observations provide $Q(rJgi)$
versus azimuthal angle (or azimuthal distance after assuming a distance to the galaxy). 
The age gradient signal can be thought of
as an asymmetric ``pulse'' (see Figure~\ref{Qmodel}), whose width is a function of
radius; the width approaches zero at CR and has larger values at smaller radii. 
The width of this pulse is analogous to a period, 
$T_Q$, of a sinusoidal signal. 
Hence, the pulse can also be characterized by its angular frequency $\omega_Q = 2\pi/T_Q$, 
and analyzed with Fourier techniques.

We compute the Fourier transform of the $Q(rJgi)$ index profiles in the annuli,

\begin{equation}
 	\hat{f}(\omega_{Q})=\int_{-\infty}^{\infty} Q(\theta) e^{-i\omega \theta}\, d\theta,
\end{equation}

\noindent where 

\begin{displaymath}
  Q(\theta) = \begin{cases} \overline{Q} \equiv \frac{1}{2\pi} \int_0^{2\pi} Q(\theta) d\theta & \mathrm{if~} \theta < 0, \\
                             Q(\theta)   & \mathrm{if~} 0 \leq \theta \leq 2\pi,            \\ 
                            \overline{Q} & \mathrm{if~} \theta > 2\pi.
              \end{cases} 
\end{displaymath}

In general, $\hat{f}(\omega_{Q})$ is complex, say, of the form $\mathrm{Re}(\omega_{Q})+i\mathrm{Im}(\omega_{Q})$.
The Fourier spectrum is obtained as $|\hat{f}(\omega_{Q})| = \sqrt{\mathrm{Re}^2+\mathrm{Im}^2}$.
We search the spectrum for frequencies with amplitudes between 0.02 and 0.06 mag. This range of amplitudes
corresponds to the ones 
achieved by the $Q(rJgi)$ index of our models,\footnote{
We consider two models in this analysis (see Figure~\ref{Qmodel}).
Model ``A'' is obtained with the SPS models only.
Model ``B'' includes both the dissolution of stellar groups~\citep{wie77} after 50 Myr, 
and stellar orbit diffusion. According to~\citet{wie77}, the diffusion of stellar
orbits can enhance the dissolution of young stellar groups by
increasing their internal velocity dispersion.}
including the photometric error of the observed $Q(rJgi)$ profiles,
$\sigma_{Q} \sim 0.05$. If more than one frequency is found, we then select the one with
the highest frequency value.\footnote{
In general, the lower frequency represents the overall shape of the
$Q(rJgi)$ profile, whereas higher frequencies correspond to small-scale structure.}
 
\begin{figure*}
\centering
\epsscale{2.0}
\plotone{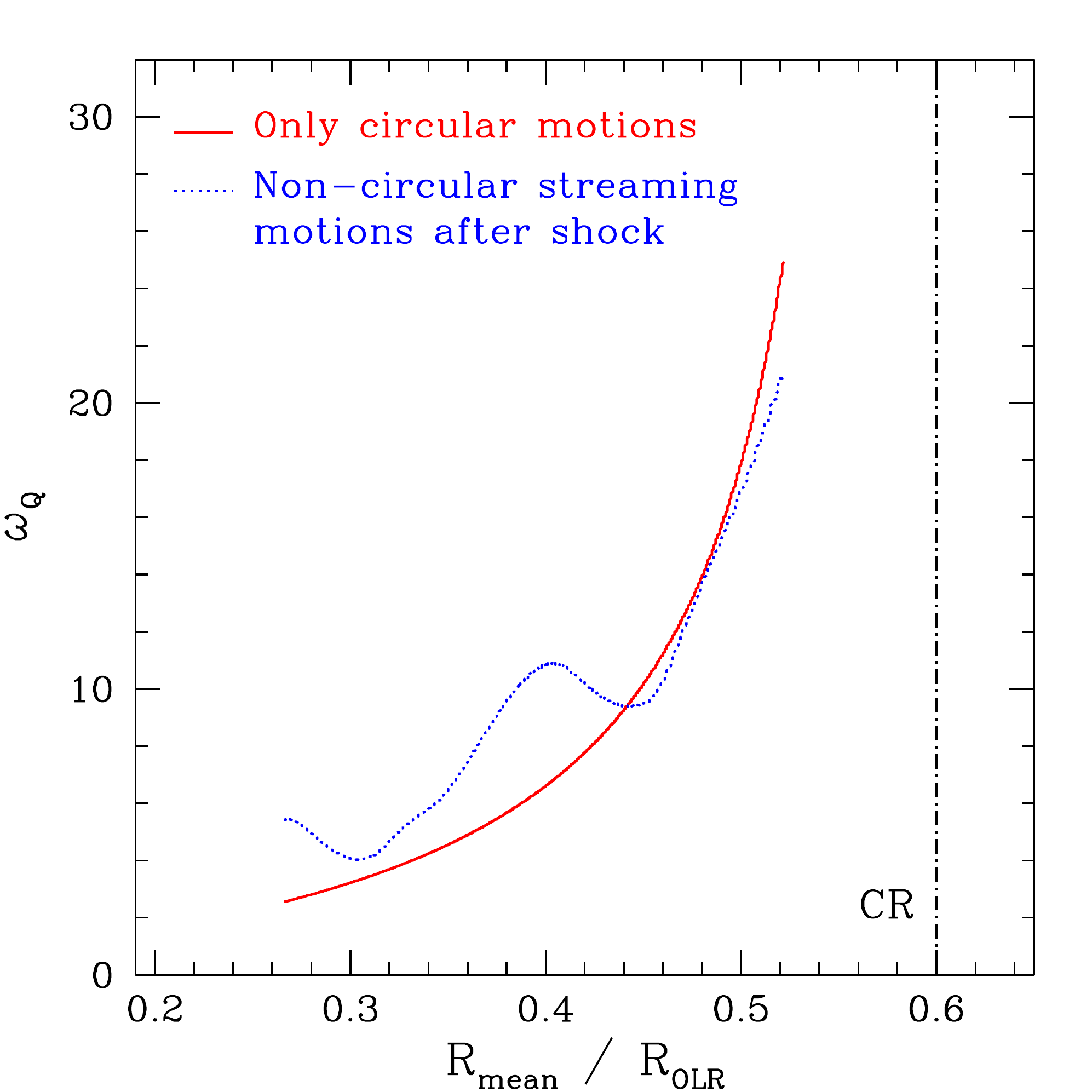}
\caption[f6]{Theoretical radial plot of $\omega_{Q}$.
Angular frequencies $\omega_{Q}$ obtained from the Fourier spectrum, 
after applying the method described in Section~\ref{Q_fourier}. 
The mean radius, $R_{\rm mean}$, has been normalized by the 
OLR radius, $R_{\rm OLR}$. 
Solid line: newborn stars follow circular trajectories.
Dotted line: results from the MHD simulation data described
in \citet{mar09b}; stars have non-circular streaming motions as a
consequence of the shock.  
Dash-dotted vertical line: corotation radius (CR).
~\label{Wplot_prediction}}
\end{figure*}

The expected radial behavior of $\omega_Q$ after applying the procedure to the MHD simulation of~\citet{mar09b}
is shown in Figure~\ref{Wplot_prediction}, with a dotted line. 
We remind the reader that the simulation has a spiral pattern 
with a constant angular speed that does not depend on radius. 
The plot also shows (solid line) the behavior of $\omega_Q$
when newborn stars move in perfectly circular orbits.
In both cases, since the width of the pulse goes to 
zero at the CR radius, its reciprocal, $\omega_Q$, diverges there; 
$R_{\rm{CR}} / R_{\rm{OLR}} \sim 0.6$.

\begin{figure*}
\centering
\epsscale{2.0}
\plotone{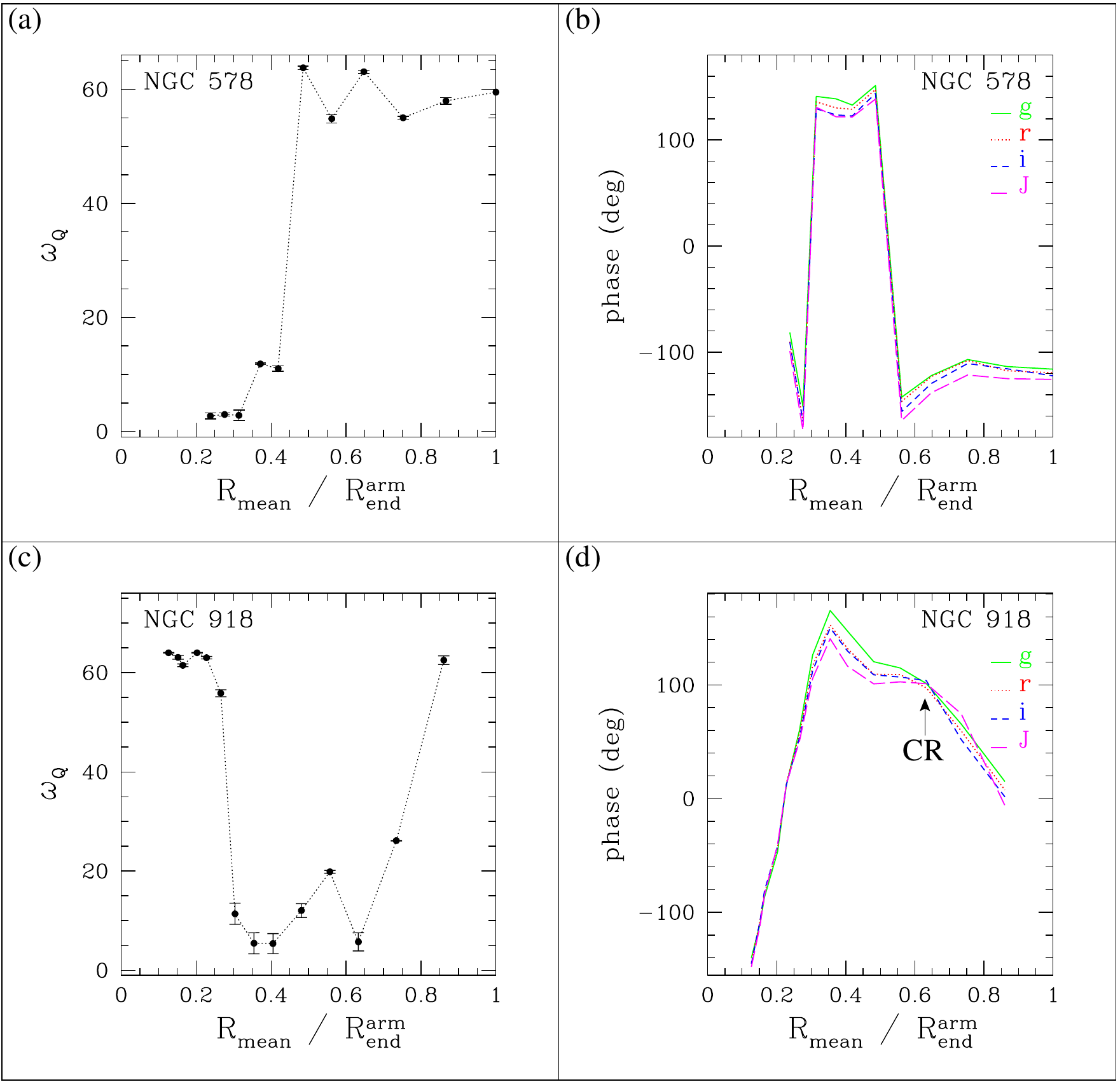}
\caption[f8]{
Panels (a) and (c): frequencies $\omega_{Q}$ vs. $R_{\rm{mean}} / R^{\rm arm}_{\rm end}$
for NGC~578, and NGC~918, respectively. Panels (b), and (d): two-armed phases (see Section~\ref{phases}),
in the $g$ (green solid line), $r$ (red dotted line),
$i$ (blue short-dashed line), and $J$ (magenta long-dashed line) bands,
vs. $R_{\rm{mean}} / R^{\rm arm}_{\rm end}$ for NGC~578 and NGC~918, respectively.
The adopted resolution in the phase analysis is~$2.8\degr$.
Corotation likely radial position is indicated for NGC~918.
~\label{WplusPHASES_1}}
\end{figure*}

\begin{figure*}
\centering
\epsscale{2.0}
\plotone{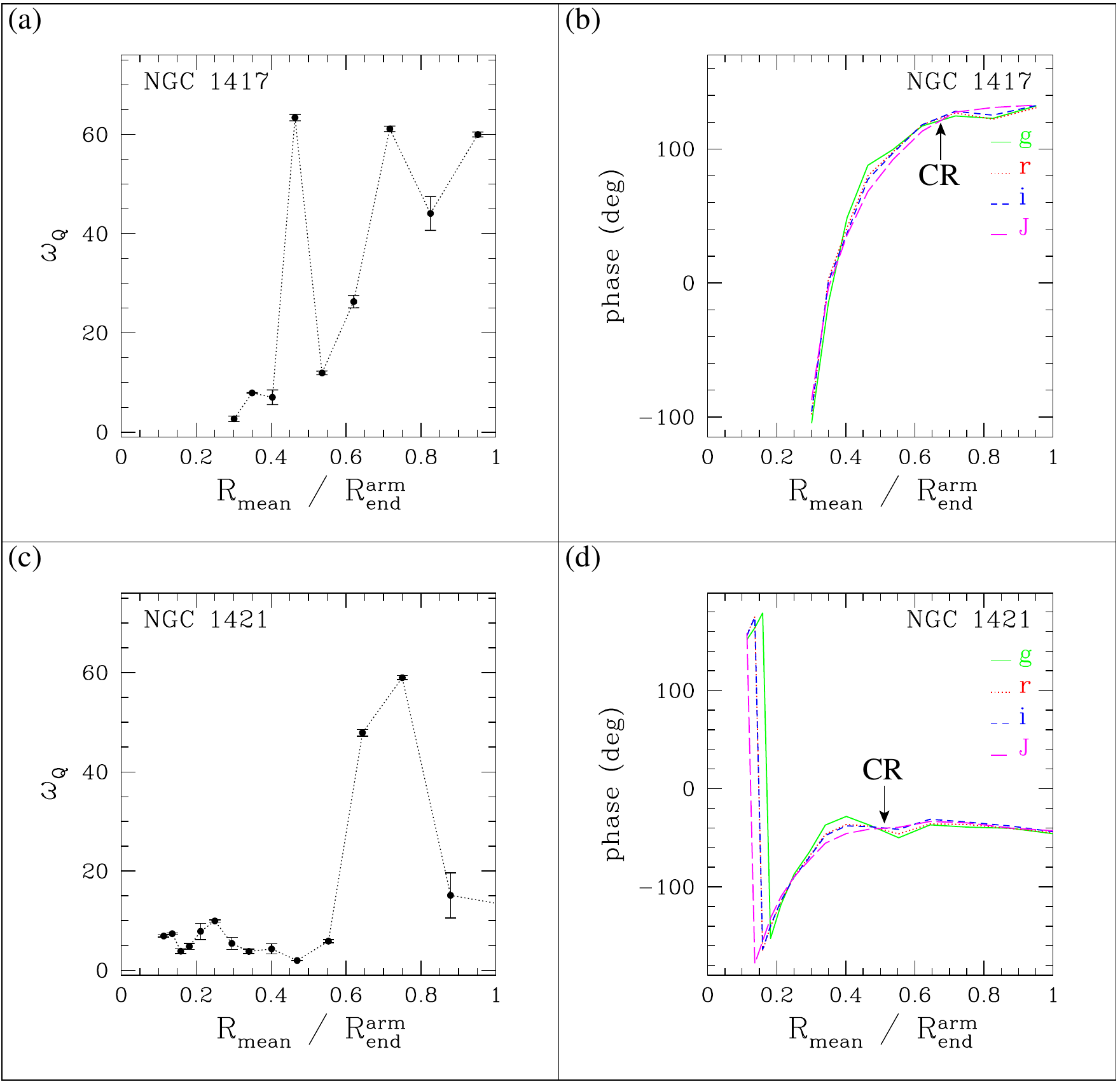}
\caption[f9]{
Panels (a), and (c): frequencies $\omega_{Q}$ vs. radius 
for NGC~1417 and NGC~1421, respectively. Panels (b) and (d): two-armed phases vs. radius 
for NGC~1417 and NGC~1421, respectively.
Corotation likely radial positions are indicated.
Symbols as in Figure~\ref{WplusPHASES_1}.
~\label{WplusPHASES_2}}
\end{figure*}

\begin{figure*}
\centering
\epsscale{2.0}
\plotone{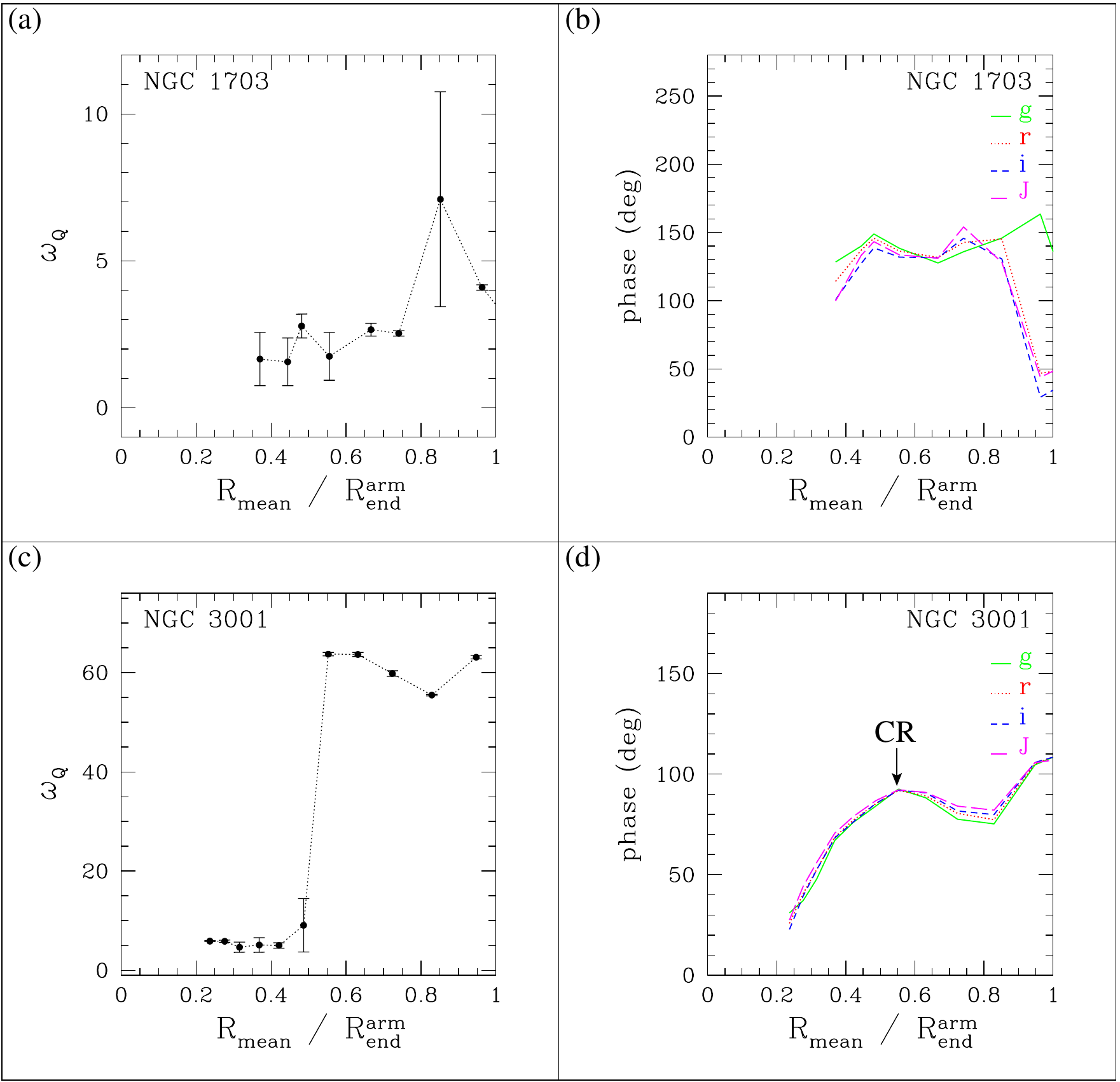}
\caption[f10]{Panels (a) and (c): frequencies $\omega_{Q}$ vs. radius 
for NGC~1703 and NGC~3001, respectively. Panels (b) and (d): two-armed phases 
vs. radius for NGC~1703 and NGC~3001, respectively.
Corotation likely radial position is indicated for NGC~3001.
Symbols as in Figure~\ref{WplusPHASES_1}.
~\label{WplusPHASES_3}}
\end{figure*}

\begin{figure*}
\centering
\epsscale{2.0}
\plotone{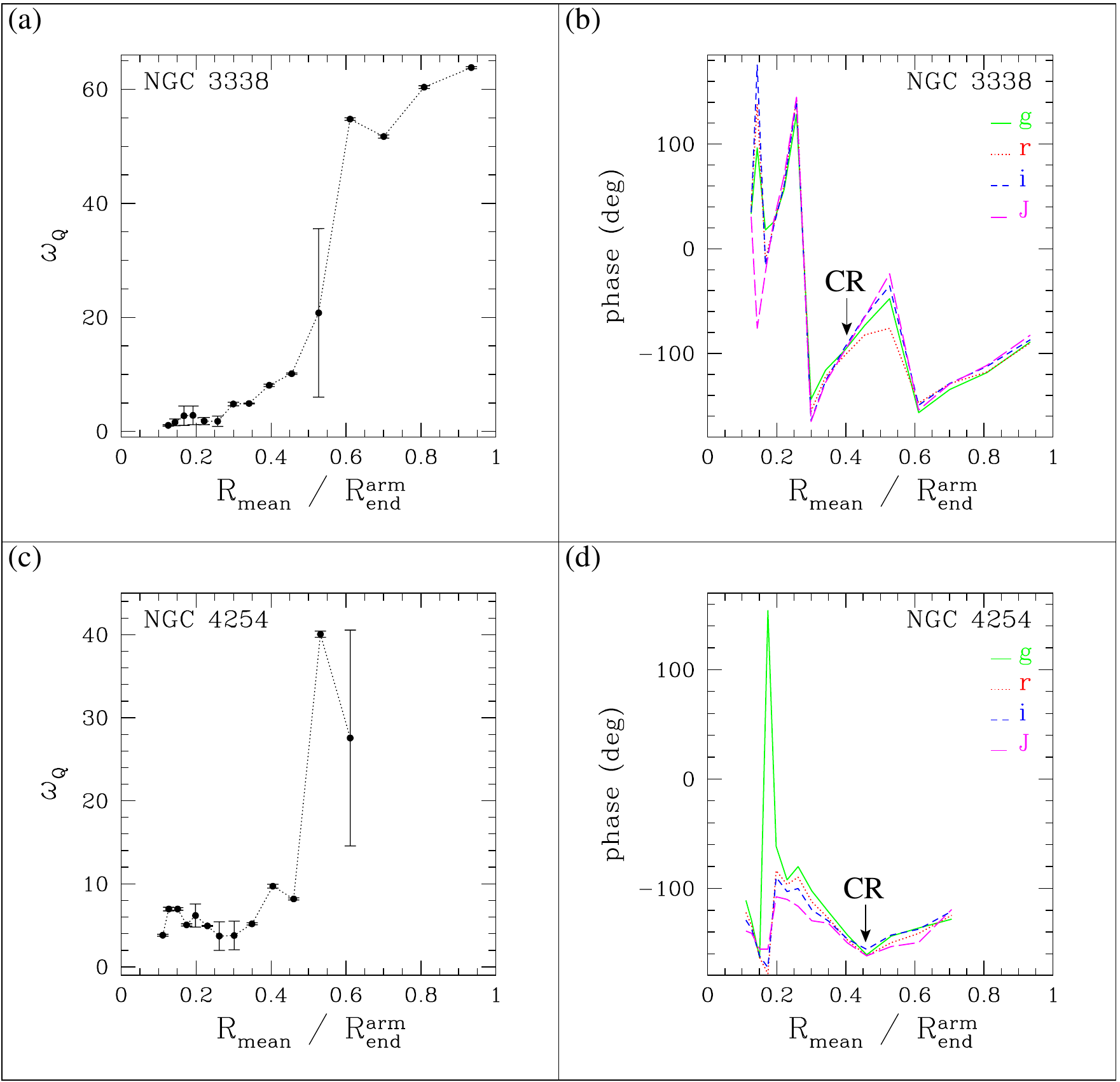}
\caption[f11]{Panels (a) and (c): frequencies $\omega_{Q}$ vs. radius 
for NGC~3338 and NGC~4254, respectively. Panels (b) and (d): two-armed phases vs. radius 
for NGC~3338 and NGC~4254, respectively.
Corotation likely radial positions are indicated.
Symbols as in Figure~\ref{WplusPHASES_1}.
~\label{WplusPHASES_4}}
\end{figure*}

\begin{figure*}
\centering
\epsscale{2.0}
\plotone{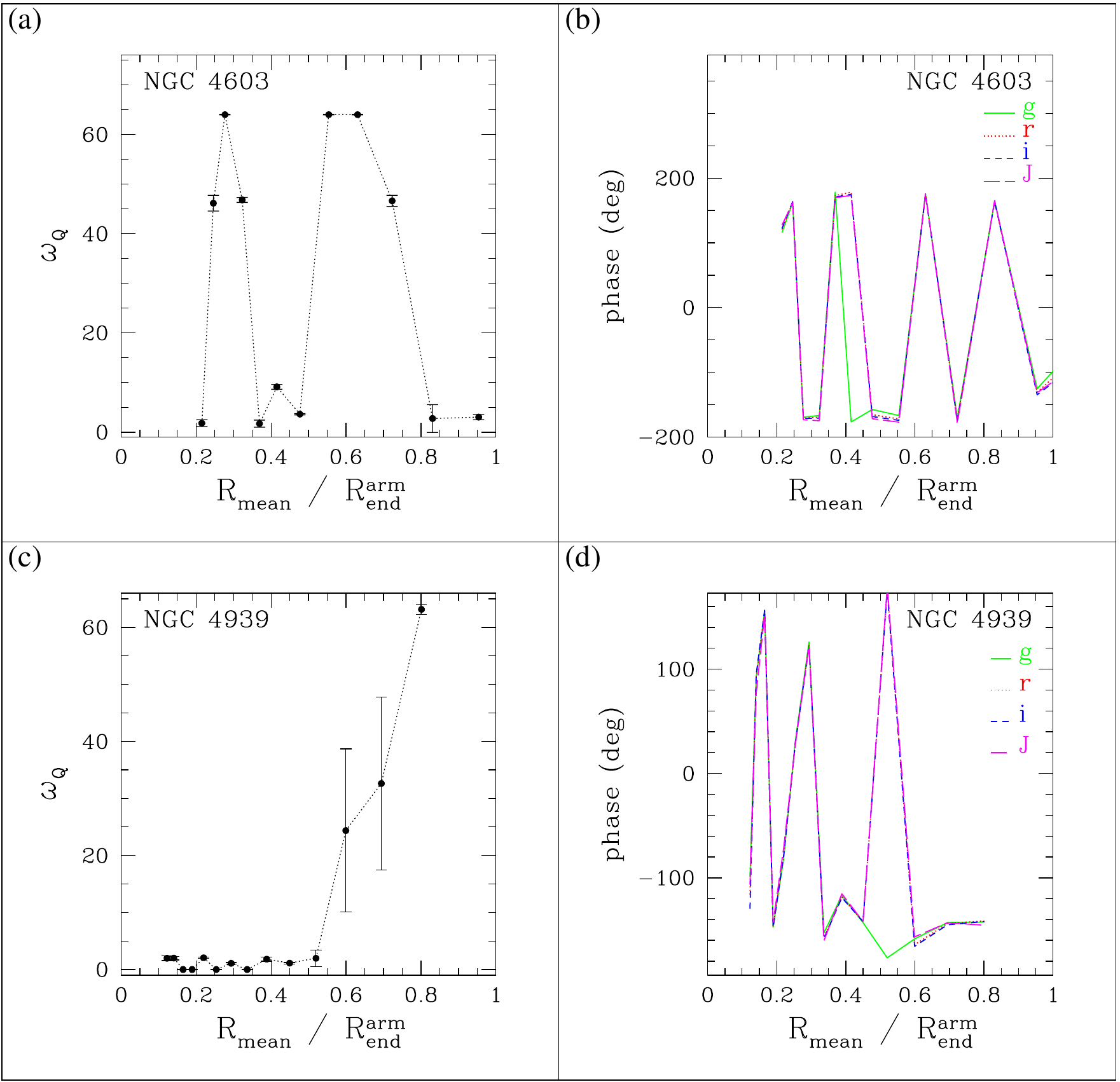}
\caption[f12]{Panels (a) and (c): frequencies $\omega_{Q}$ vs. radius 
for NGC~4603 and NGC~4939, respectively. Panels (b) and (d): two-armed phases vs. radius 
for NGC~4603 and NGC~4939, respectively.
Symbols as in Figure~\ref{WplusPHASES_1}.
~\label{WplusPHASES_5}}
\end{figure*}

\begin{figure*}
\centering
\epsscale{2.0}
\plotone{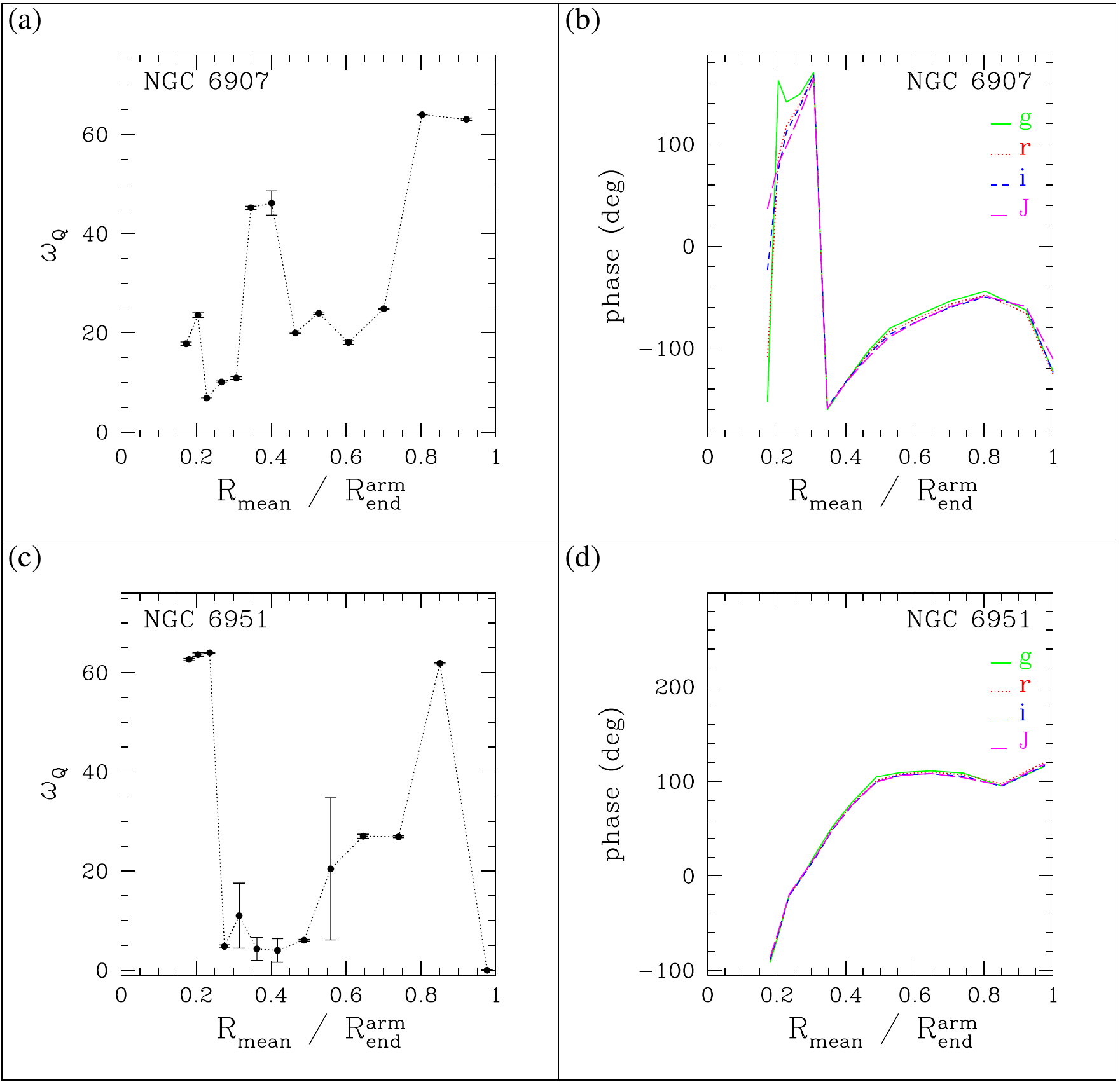}
\caption[f13]{Panels (a) and (c): frequencies $\omega_{Q}$ vs. radius 
for NGC~6907 and NGC~6951, respectively. Panels (b) and (d): two-armed phases vs. radius 
for NGC~6907 and NGC~6951, respectively.
Symbols as in Figure~\ref{WplusPHASES_1}.
~\label{WplusPHASES_6}}
\end{figure*}

\begin{figure*}
\centering
\epsscale{2.0}
\plotone{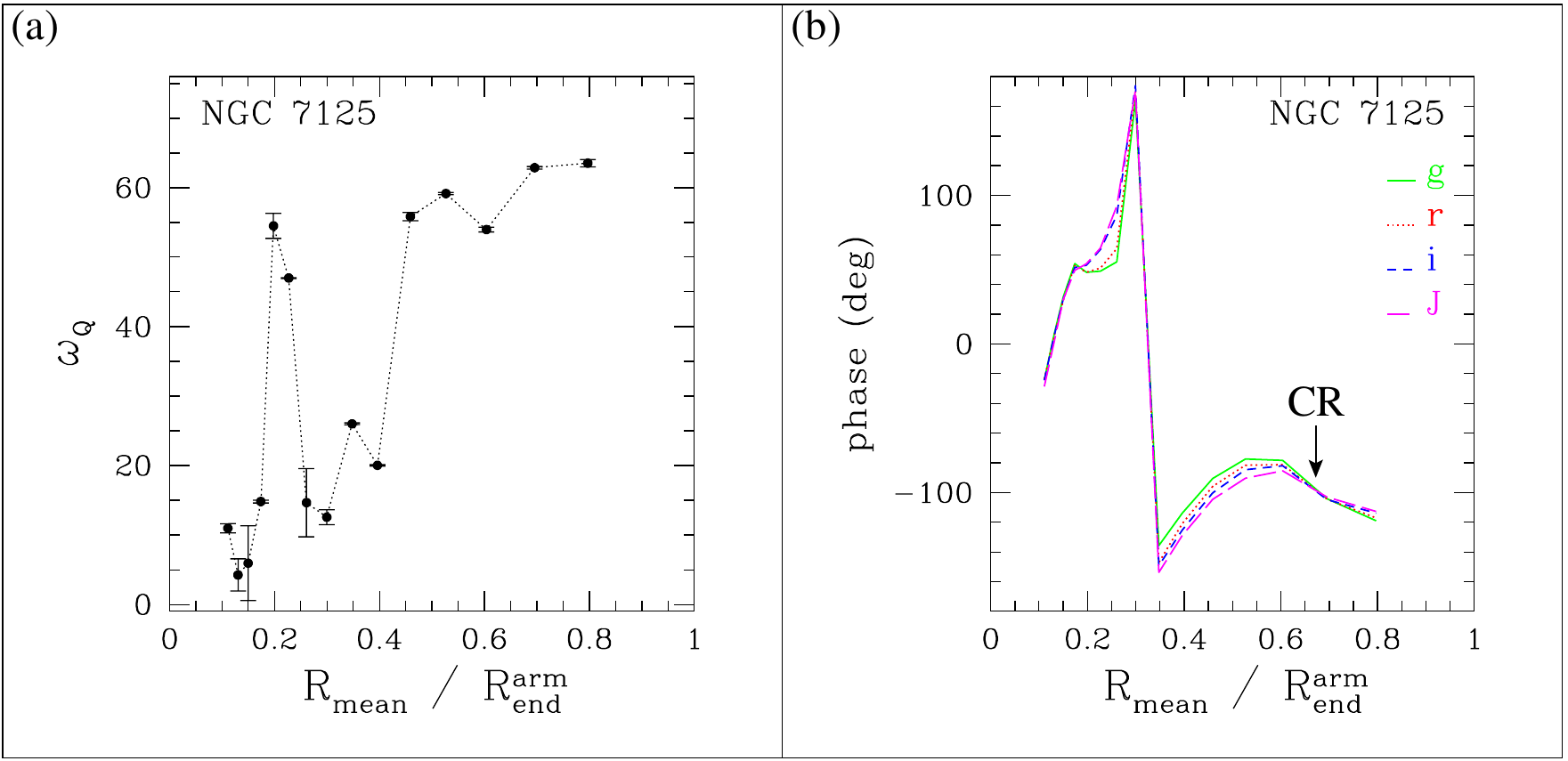}
\caption[f14]{Panel (a): frequencies $\omega_{Q}$ vs. radius 
for NGC~7125. Panel (b): two-armed phases vs. radius for NGC~7125.
Corotation likely radial position is indicated.
Symbols as in Figure~\ref{WplusPHASES_1}.
~\label{WplusPHASES_7}}
\end{figure*}

Figures~\ref{WplusPHASES_1} through~\ref{WplusPHASES_7}, left panels, show the results
of this analysis when applied to our sample of objects. Data $\omega_{Q}$ values are graphed versus 
$R_{\rm{mean}} / R^{\rm arm}_{\rm end}$, where $R_{\rm{mean}}$ is the mean galactocentric radius, 
and $R^{\rm arm}_{\rm end}$ is the
spiral maximum radial extent. The latter is determined visually in the NIR band 
indicated in Table~\ref{tbl-armend}.


\begin{deluxetable}{llr}
\tabletypesize{\scriptsize}
\tablecaption{Spiral Radial Maximum Extent~\label{tbl-armend}}
\tablewidth{0pt}
\tablehead{
\colhead{Galaxy}
& \colhead{$R^{\rm arm}_{\rm end}$ (arcsec)}
& \colhead{$R^{\rm arm}_{\rm end}$ (kpc)}
}

\startdata

NGC~578  & ~95.7 $\pm$  2.9     ($K_s$)     &   10.5    $\pm$   0.9 \\
NGC~918  & ~75.4 $\pm$  5.8     ($K^\prime$)&   7.9     $\pm$   0.7 \\
NGC~1417 & ~60.0 $\pm$  5.0     ($J$)       &   16.6    $\pm$   1.4 \\
NGC~1421 & ~92.8 $\pm$  2.9     ($K_s$)     &   13.2    $\pm$   1.1 \\
NGC~1703 & ~27.5 $\pm$	1.4 	($H$)	    &	2.7	    $\pm$	0.3 \\
NGC~3001 & ~44.0 $\pm$	2.8	    ($H$)	    &	7.6  	$\pm$	0.8 \\
NGC~3338 & 145.0 $\pm$	5.0	    ($K_{s}$)   &	16.7	$\pm$	1.6 \\
NGC~4254 & 157.5 $\pm$  7.5     ($K_s$)     &   12.6    $\pm$   0.8 \\ 
NGC~4603 & ~45.4 $\pm$	1.4	    ($H$)	    &	6.5	    $\pm$	0.6 \\
NGC~4939 & 145.0 $\pm$  5.0     ($K_s$)     &   32.7    $\pm$   2.8 \\
NGC~6907 & ~55.1 $\pm$	2.9	    ($K_{s}$)   &	12.7	$\pm$	1.3 \\
NGC~6951 & 105.0 $\pm$  5.0     ($K_s$)     &   12.7    $\pm$   1.1 \\
NGC~7125 & ~95.7 $\pm$  2.9     ($J$)       &   20.7    $\pm$   1.8 \\


\enddata

\tablecomments{
Columns 2 and 3: spiral radial maximum extent (determined visually), in arcsec and kpc, respectively, and bandpass used to determine it.
}

\end{deluxetable}


\subsection{Azimuthal Phases of the $g$, $r$, $i$, and $J$ Bands}~\label{phases}

If DWs indeed produce shocks and
induce star formation, we expect increasingly older stellar populations
(with intensity peaks at longer wavelengths) to be 
sequentially located downstream (in the direction of rotation) from the shock position 
within CR, and to be sequentially placed upstream (opposite rotation) outside CR.
No gradients should be observed at CR. 
We use also Fourier techniques to analyze  
our sample in a systematic manner, at all galactic radii, in
search of this effect. 

We adopt the method of~\citet{pue97}
to determine the phases (basically, the conjugate of the azimuthal positions) of stellar populations of different ages, as
traced in the $g$, $r$, $i$, and $J$ bands. 
The method is based on computing the Fourier transform of the form

\begin{equation}
    \hat{f}(m)=\int_{-\pi}^{\pi} I_{R}(\theta) e^{-im \theta}\, d\theta,
\end{equation}

\noindent where $I_{R}$ is the intensity of radiation, and with phase

\begin{equation}
    \Phi = \tan^{-1} \left\{ \frac{\mathrm{Re}[\hat{f}(m)]}{\mathrm{Im}[\hat{f}(m)]} \right\}.    
\end{equation}

\noindent
Given that we have selected objects with a strong two-armed pattern, $m=2$.
This means that 2$\pi$ radians include two periods of the analyzed signal, i.e., there is 
a $\pi$ radian symmetry. Although this may not be exactly true for real arms in spirals (P.~Grosb{\o}l 2012, private
communication), it is a very good approximation for our purposes.
We analyze the
same concentric annuli in the ``straightened'' images as before.

\begin{figure*}
\centering
\epsscale{2.0}
\plotone{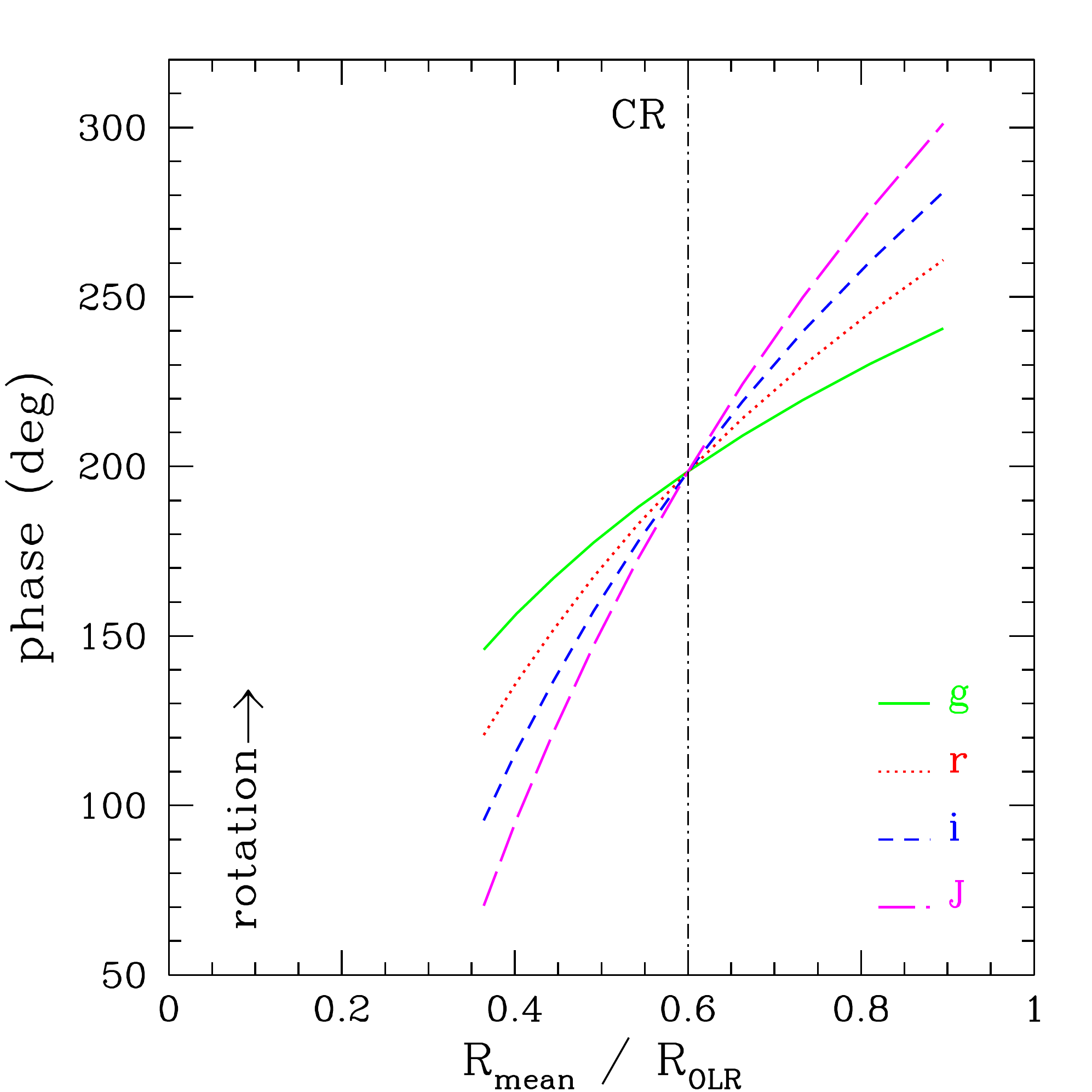}
\caption[f7]{Schematic of the theoretical phase values (in degrees) at
different wavelengths vs. radius for a density wave with $m=2$.
The mean radius, $R_{\rm mean}$, has been normalized by the OLR radius, $R_{\rm OLR}$.
Green solid line: $g$ band ($\lambda_{\mathrm{eff}} = 5000\ {\rm \AA}$); red dotted line: $r$ band ($\lambda_{\mathrm{eff}} = 6800\ {\rm \AA}$);
blue short-dashed line: $i$ band ($\lambda_{\mathrm{eff}} = 7800\ {\rm \AA}$); magenta long-dashed line: $J$ band ($\lambda_{\mathrm{eff}} = 12500\ {\rm \AA}$).
Dash-dotted vertical line: corotation radius (CR).
~\label{PHASEplot_prediction}}
\end{figure*}

For the data analysis we assume all spirals in our sample trail,
and hence that the spiral opens in the direction opposite rotation.
Furthermore, we adopt the convention that the angular
coordinate increases with rotation,
regardless of whether the arms have an ``S'' or a ``Z'' on-the-sky view.
Under these premises, 
for a wave of the form $\cos(\theta+\Phi)$,
where $\Phi$ is the phase, $\Phi$ should decrease with wavelength
inside CR and increase outside CR. There should be
no phase difference with wavelength at CR (see Figure~\ref{PHASEplot_prediction}). 
The results of this exercise are shown in Figures~\ref{WplusPHASES_1} through~\ref{WplusPHASES_7},
right panels. 

\section{Discussion}

For seven objects (NGC~918, NGC~1417, NGC~1421, NGC~3001, NGC~3338, NGC~4254, and NGC~7125, 
see Table~\ref{tbl-CRll}), we find similarities between the theoretical expectations
and the observations, regarding both 
the radial distribution of $\omega_Q$ and the radial run of intensity phases with wavelength,
at least for some range of radii.
However, all the objects also show discrepancies. 
For example, in NGC~918, NGC~1417, NGC~1421, NGC~3338, NGC~4254, and NGC~7125,
the phases of all four passbands overlap between $\approx$ 0.2 and 0.4 $R_{\rm mean}/R^{\rm arm}_{\rm end}$,
but then behave in accordance to theory up to what is likely the CR 
radius, as judged by the position where $\omega_Q$ diverges and where, again, phases 
are the same for all wavelengths.
In NGC~918 and NGC~4254, phase differences agree with theory within
CR, but then do not change direction. 
This could be due to a weaker spiral potential and, hence, weaker shocks at radii beyond CR.
Conversely, in NGC~3001 the phases show the expected behavior only outside the CR position,
if it is indeed located where the $\omega_{Q}$ plot shows a clear divergence. 
The intensity phases in NGC~1703 (which may be an SB type galaxy) behave as predicted by theory, 
but $\omega_{Q}$ does not diverge at the CR position expected from the phases.

From this analysis, there is not even one object for which both diagnostics 
agree perfectly with theory at all radii. These findings confirm more objectively
a previous result from visual inspection: 
age/color textbook gradients that run coherently along
entire arms and change direction at CR are 
very hard to detect in practice. Possible causes
may be a low S/N of the data, or masking of
the gradients by the intense and clumpy (continuum and line) radiation from massive star clusters. 
In the case of the intensity phases method, dusty environments can attenuate
the shorter wavelengths, hindering the detection of an ordered age sequence.\footnote{
This was the original motivation behind the development of the reddening-insensitive index
$Q(rJgi)$.}
It is also possible that the gradients are perturbed by
the ``infant mortality'' of star clusters~\citep{lad03},
or by the destruction of an ordered age front by 
supernovae shocks and stellar winds \citep{mu76,ger78}.
On the other hand, even in the presence of DWs, 
it is possible that no gradients are produced. For example, 
there may be inadequate physical conditions for the creation of the
gradients, related to the strengths of the shock or of the magnetic field \citep{efr10}.
There is, of course, the possibility that a real constant pattern speed for all radii
does not exist, i.e., that the spiral structure is not explained by the DW theory. 
Our subsample, however, was selected explicitly to minimize this eventuality. 
Furthermore, we had previously established the link of the gradients in the sample, 
taken as an ensemble, to disk dynamics. As confirmed by the comparison with 
an MHD simulation with a constant spiral pattern speed,  
an apparent radial dependency of $\Omega_{\rm p}$ is caused by
neglecting non-circular streaming motions in the analysis of the gradients.

Conversely, there could be 
color gradients that are not linked to the dynamics of the disk. 
One speculative scenario is provided by the IGIMF
(the Integrated Galaxial IGIMF) theory of~\citet{pfl08} and~\citet{pfla09}. This theory is able to explain the
H$\alpha$ cutoff in galaxy disks without a corresponding cutoff
in star formation (as deduced from the non-ionizing far-ultraviolet flux
given out by B-type stars). The IGIMF theory posits that maximum star cluster mass 
depends on gas surface density (or, equivalently, on star formation rate),\footnote{This has 
been corroborated by \citet{gon12} in
M33.} and that maximum stellar mass 
depends on cluster mass. If this is indeed the case,
one could obtain a color gradient from coeval star clusters across a region with a gas surface density gradient. 

At any rate, a statistical analysis, tied to spiral dynamics,
of color gradients in several objects is necessary to identify the gradients 
that are linked to DWs. This is even more important since
by general rule gradients are painstakingly
identified only in selected regions and not along entire arms.

It is, however, important to stress that our new results imply that 
long-lived modes seem to have produced gradients that run for a
sizeable part of a spiral arm in at least $50\%$ of the analyzed objects. 
This is, in itself, remarkable.

For the rest of the objects 
(NGC~578, NGC~1703, NGC~4603, NGC~4939, NGC~6907, and NGC~6951), the Fourier methods
failed to detect signs of gradients besides the ones that had already
been identified by eye.
It is interesting to point out, though, that NGC~4603 shows 
a multi-armed structure in the optical and NIR bands, although it was kept in 
the sample because of its two-armed structure in the mass map. 
For NGC~6951 (type SAB in the RC3) and NGC~6907 (type SB in the RC3),
the CR radius may lie around $R^{\rm arm}_{\rm end}~\sim 0.8$.
This is possibly an indication of color gradients within the bar region,
as already found in~\citet{mar11}.

We find no relation between Hubble type and success rate in our Fourier tests. 


\begin{deluxetable}{lccr}
\tabletypesize{\scriptsize}
\tablecaption{Corotation Positions ($R_{\rm CR}$)~\label{tbl-CRll}}
\tablewidth{0pt}
\tablehead{
\colhead{Galaxy}
& \colhead{$R_{\rm{mean}} / R^{\rm arm}_{\rm end}$}
& \colhead{$R_{\rm CR}$ (arcsec)}
& \colhead{$R_{\rm CR}$ (kpc)}
}

\startdata

NGC~918	         &  0.63 &  	$47.5\pm3.7$  &		$5.0 \pm0.4$   \\
NGC~1417	 &  0.67 &  	$40.2\pm3.4$  &		$11.1\pm0.9$   \\
NGC~1421	 &  0.49 &  	$45.5\pm1.4$  &		$6.5 \pm0.5$   \\
NGC~3001	 &  0.55 &  	$24.2\pm1.5$  &		$4.2 \pm0.4$   \\
NGC~3338	 &  0.38 &  	$55.1\pm1.9$  &		$6.3 \pm0.6$   \\
NGC~4254	 &  0.46 &	$72.5\pm3.5$  &	        $5.8 \pm0.4$   \\
NGC~7125	 &  0.67 &  	$64.1\pm1.9$  &		$13.9\pm1.2$   \\

\enddata

\tablecomments{
Column 1: object with signatures of a radially extended azimuthal gradient.
Column 2: corotation likely position, $R_{\rm{mean}} / R^{\rm arm}_{\rm end}$. The mean value for the six objects is $R_{\rm{mean}} / R^{\rm arm}_{\rm end} = 0.55$.
Columns 3, and 4: corotation radius, $R_{\rm CR}$, in arcseconds and kiloparsecs, respectively.
}

\end{deluxetable}


\subsection{Pitch Angles Test}~\label{pitch_angles}

As a complementary test of DW presence, we examine
the pitch angles of the spiral arms in different wavebands ($g$, $r$, $i$, and $J$), and
in the ``resolved mass maps'' previously obtained with the method of~\citet{zib09}.
As already mentioned in the introduction, according to~\citet{efr10},
age gradients are observed across arm segments 
that fulfill the conditions for a strong shock to occur. 
These segments seem to have an irregular
magnetic field, while the regions with a regular magnetic field
harbor instead weak shocks, and chains of star complexes with a nearly regular spacing. 
Strong shocks are associated with a larger pitch angle of the spiral segment.

Another prediction involving pitch angles in the DW theory paradigm
comes from the local dispersion relation. At a fixed mass surface density, 
the pitch angle increases proportionally to the square of the 
velocity dispersion~\citep[see, e.g.,][]{hoz03,atha10,mar12}, that is,
with the age of the stellar population and,
hence, with image wavelength in the optical and NIR. 

Pitch angles were measured with Fourier techniques~\citep{con82,con88,pue92,sar94,dav12,sav12}, in
the radial range where the spiral perturbation seems more prominent (see Table~\ref{tbl-pitchs}). This method assumes a
logarithmic geometry for the spirals, although this might not be the case for all objects at all radii~\citep{rin09}.

\begin{figure*}
\centering
\epsscale{2.0}
\plotone{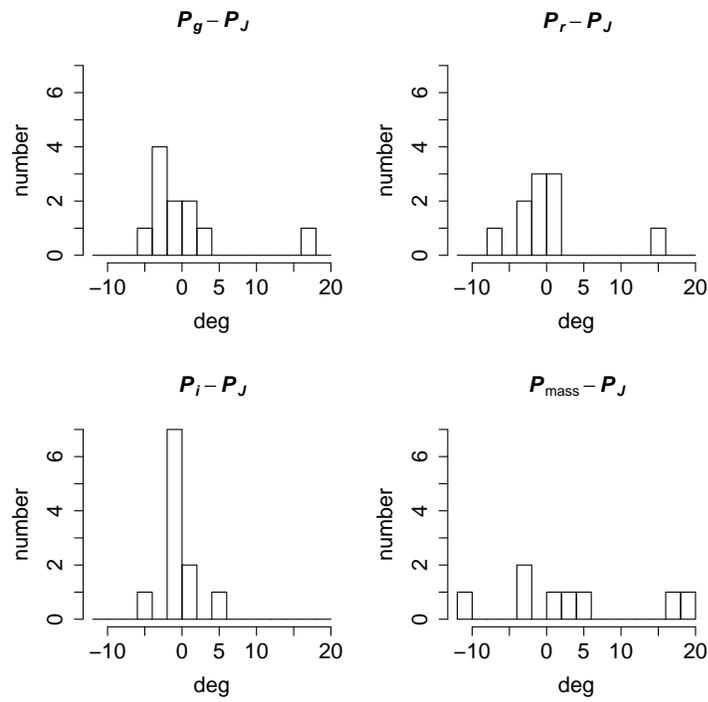}
\caption[f15]{Histograms of the pitch angle differences between the optical bands $g$ (top left),
$r$ (top right), $i$ (bottom left), the ``mass map'' (bottom right), and the NIR $J$ band.
~\label{HISTO_J_gri}}
\end{figure*}

Figure~\ref{HISTO_J_gri} shows histograms of $P_x$ - $P_J$, 
where $P$ is the pitch angle and $x$ can be $g$, $r$, $i$, or ``mass'', depending on
the image where $P$ has been measured.\footnote{
Absolute values from Table~\ref{tbl-pitchs} were adopted for analysis.
However, $P_x$ values were discarded if their sign 
does not agree with that of $P_J$. (Signs indicate whether the galaxy has an
``S'' or a ``Z'' on-the-sky view.)
The pitch angles measured for NGC~578 with the Fourier method were also rejected,
since their nominal sign is at odds with the observed on-the-sky view.}
Negative values of the differences indicate that the pitch angle is larger in the NIR
than in the optical. In the histograms of $P_{g}-P_{J}$, $P_{r}-P_{J}$, and $P_{i}-P_{J}$,
there are indeed more objects with negative differences. 
The median values of each histogram are, respectively, -0.9$\degr$, -1.1$\degr$, and 0.0$\degr$.
This result is in agreement with~\citet{gro98}, who find 
tighter arms in bluer colors in images of four spirals, suggesting the presence of DWs.

Interestingly, the median value of the $P_{\rm{mass}}-P_{J}$ histogram is 2.2$\degr$, so that
the mass arms are statistically more open than the arms in NIR light.  
An inspection of the mass maps indicates that the spiral arms loci in the NIR images do not always coincide
with the location of the arms in the mass maps~(see Figures~\ref{J_vs_Mass} and~\ref{J_vs_Mass_cont}),
a fact that can be explained by the contribution of young stars to the NIR light.

\begin{figure*}
\centering
\epsscale{2.0}
\plotone{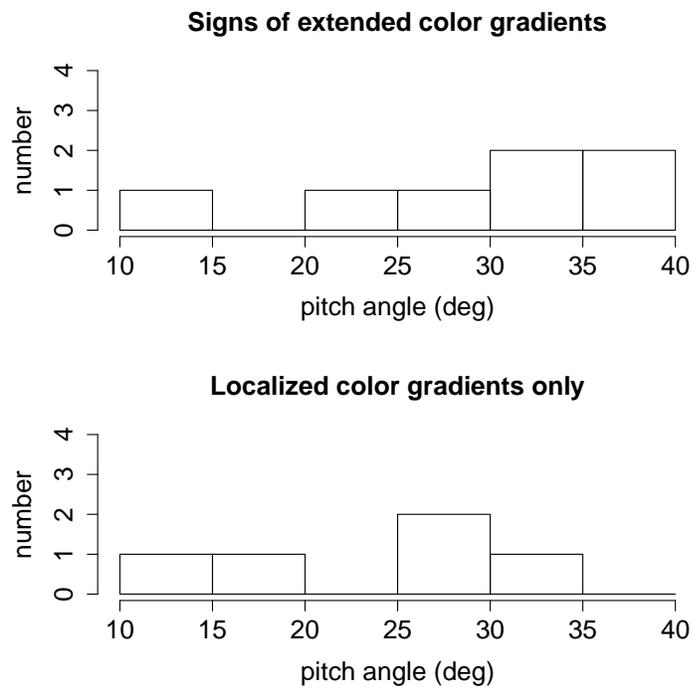}
\caption[f16]{Histograms of the median pitch angle values. Top: objects with signs of
widespread color gradients; 
bottom: objects with localized color gradients only. 
~\label{HISTO_longlived}}
\end{figure*}

Finally, we compare the median pitch angle
distribution in those galaxies that we have classified as having radially extended azimuthal 
color gradients, against the distribution in the
galaxies where we have detected localized gradients only. 
Figure~\ref{HISTO_longlived} displays the pitch angle histograms for
each data set. The statistical significance is low, due to the small
number of objects, but the objects with signs of widespread color gradients 
may have more open spiral arms (larger pitch angles).


\begin{deluxetable}{crrrrrrc}
\tabletypesize{\scriptsize}
\tablecaption{Pitch Angle Values~\label{tbl-pitchs}}
\tablewidth{0pt}
\tablehead{
\colhead{Galaxy}
& \colhead{$\Delta{R}$(arcsec)}
& \colhead{$P_{g}$}
& \colhead{$P_{r}$}
& \colhead{$P_{i}$}
& \colhead{$P_{J}$}
& \colhead{$P_{\rm{mass}}$}
& \colhead{``Median''}
}

\startdata

NGC~578    &  (20.3-63.8)  &	-34.70  ($m$=2)    &  -35.72   ($m$=2)    & -37.92   ($m$=2)    &    -41.68   ($m$=2)    &  -49.45   ($m$=2) &   40.97\tablenotemark{a}  \\
NGC~918	   &  (18.6-57.2)  &     20.60  ($m$=2)    &   21.77   ($m$=2)    &  22.63   ($m$=2)    &     23.54   ($m$=2)    &   29.42   ($m$=2) &   22.20   \\
NGC~1417   &  (20.0-60.0)  &	 36.82  ($m$=2)    &   33.76   ($m$=2)    &  36.82   ($m$=2)    &     41.71   ($m$=2)    &   39.13   ($m$=2) &   36.82   \\ 
NGC~1421   &  (34.4-91.6)  &	 36.17  ($m$=2)    &   37.17   ($m$=2)    &  38.21   ($m$=2)    &     39.31   ($m$=2)    &  -81.66   ($m$=2) &   37.69   \\   
NGC~1703   &  (12.4-27.5)  &	 16.55  ($m$=2)    &   16.29   ($m$=2)    &  15.79   ($m$=2)    &    -36.16   ($m$=1)    &  -55.62   ($m$=2) &   16.04   \\
NGC~3001   &  (19.2-44.0)  &	 31.55  ($m$=2)    &   31.55   ($m$=2)    &  32.45   ($m$=2)    &     32.45   ($m$=2)    &   34.40   ($m$=2) &   32.00   \\
NGC~3338   &  (45.0-145.0) &	-13.43  ($m$=2)    &  -13.60   ($m$=2)    & -13.77   ($m$=2)    &    -13.77   ($m$=2)    &   83.87   ($m$=2) &  -13.68   \\
NGC~4254   &  (30.0-157.5) &	-29.30  ($m$=3)    &   18.13   ($m$=1)    & -29.30   ($m$=3)    &    -29.30   ($m$=3)    &   22.62   ($m$=1) &  -29.30   \\
NGC~4603   &  (13.8-45.4)  &	 25.61  ($m$=2)    &   26.87   ($m$=2)    &  27.54   ($m$=2)    &     28.25   ($m$=2)    &   30.59   ($m$=2) &   27.20   \\
NGC~4939   &  (50.0-145.0) &	 10.96  ($m$=2)    &   11.07   ($m$=2)    &  11.07   ($m$=2)    &     10.74   ($m$=2)    &   29.89   ($m$=3) &   11.02   \\
NGC~6907   &  (34.8-55.1)  &	 24.89  ($m$=2)    &   24.34   ($m$=2)    &  27.34   ($m$=2)    &     27.34   ($m$=2)    &   45.15   ($m$=2) &   26.12   \\
NGC~6951   &  (45.0-105.0) &	-31.48  ($m$=2)    &  -33.26   ($m$=2)    & -32.35   ($m$=2)    &    -34.23   ($m$=2)    &  -36.30   ($m$=2) &  -32.80   \\
NGC~7125   &  (20.3-95.7)  &	-23.88  ($m$=2)    &  -26.76   ($m$=2)    & -35.99   ($m$=2)    &    -40.84   ($m$=2)    &  -52.36   ($m$=2) &  -31.38   \\

\enddata

\tablecomments{
Column 1: analyzed object.
Column 2: radial ranges where the spiral seems more prominent, in arcseconds.
Columns 3-6: pitch angles, in degrees, for the $g$, $r$, $i$, and $J$ bands, respectively.
Negative values indicate a ``Z'' on-the-sky view for the object, while positive values indicate an ``S'' on-the-sky view.
Absolute values were adopted for analysis.
The mean error is~$\sim 1 \degr$ (without taking into account the error due to the deprojection parameters).
The dominant Fourier mode, $m$, is indicated in parenthesis.
Column 7: pitch angle, in degrees, for the resolved maps of stellar mass surface density~\citep[``mass map'',][]{zib09}.  
Column 8: median pitch angle, in degrees, adopted to ``straighten'' the spiral arms (see Section~\ref{Q_fourier}).
We define the median pitch angle for each object by arranging the $P_{g}$, $P_{r}$, $P_{i}$, and $P_{J}$ values
from lowest to highest, and averaging together the two middle values.
}

\tablenotetext{a}{Computed by fitting the arms' slope in the $\theta$ vs. ln $R$ polar image~\citep[see, e.g.,][]{mar12}.
The Fourier method yields a ``Z'' on-the-sky view, but the object actually has an ``S'' view.}

\end{deluxetable}


\section{Conclusions}~\label{conclusions}

From a Fourier analysis of 13 objects, with a spiral arm structure
that can be likely explained by DW theory, 
we find that $\sim 50\%$ of them show evidence of
color gradients that run for at least a range of radii.
For the remaining objects, the evidence of extended gradients is insufficient.
This notwithstanding, our previous studies~\citep{gon96,mar09a,mar09b,mar11} suggest that azimuthal
age/color gradients (or candidates) can be found for most galaxies, at least in some regions of the spiral arms.
A relation of such very localized gradients with the dynamics of the disk
has been previously established by the comparison with an MHD simulation with a constant
pattern speed at all radii~\citep{mar09b}.

Our results are consistent with other observational studies
of age patterns (or offsets) across spiral arms~\citep{efr85,tam08,egu09,gro09,san11}.
By means of an H$\alpha$ to far-UV flux ratio method,~\citet{san11} found age gradients
across the spiral arms of the grand-design spirals M74 and M100 (NGC~628 and NGC~4321, respectively).
Likewise, the ``breaks'' in the radial metallicity distribution found near CR
in a sample of $\approx$ 20 galaxies by ~\citet{sca11} and~\citet{sca13} imply
that spiral arms must be long-lived structures. A dominant pattern speed
must exist with a unique CR radius, or otherwise any discontinuities in the
radial metallicity profiles would be smoothed out~\citep{sca13}.

On the other hand, simulations have shown mutually exclusive results.
Long-lived quasi-steady features have been obtained for some
models~\citep[e.g.,][]{don94,zha98}. Other works, however~\citep[e.g., ][]{sel11,wada11,fuj11},
have found a lack of a persistent pattern, and instead
recurrent short-lived transient spirals only.

Any model that intends to explain spiral structure has to avoid
the winding dilemma, that ultimately led to the idea of stationary
DWs with a fixed pattern speed for all radii.
For this same reason, the observational results of~\citet{mer06} and~\citet{mei09} for some objects, where
$\Omega_{\mathrm{p}}$ decreases with radius, imply the existence of transient spiral modes
that emerge and disappear at different stages in the disk evolution. Such 
transient arms cannot produce stellar age/color gradients, though.

A theory that reconciles the existence of age/color gradients with 
a radially varying pattern speed is still missing. However,
as argued in Section~\ref{OmegaVar}, the observation of a pattern speed that
varies with radius may be an artifact produced by the non-circular streaming motions of the 
stars newly born in shocked material. This effect can explain 
the coexistence of azimuthal age/color gradients across spiral arms, and of a spiral 
pattern speed $\Omega_{\rm p}$ that seems to vary with radius in the same way as the 
orbital frequency $\Omega$. 

\acknowledgments

We acknowledge Dr. Yuri Efremov, the referee, for his comments and suggestions
that have greatly improved this paper. Special thanks go to Ivanio Puerari, for helpful discussions about Fourier
analysis of spiral galaxies. E.M.G. acknowledges a CONACYT grant for a postdoctoral fellowship at INAOE,
and former postdoctoral financial support from UNAM (DGAPA), M\'exico. R.A.G.L. is also
grateful for financial support from both CONACYT and DGAPA-UNAM.


\begin{thebibliography}{}

\bibitem[Arzoumanian et al.(2011)]{arz11} Arzoumanian, D., Andr{\'e}, P., Didelon, P., et al.\ 2011, \aap, 529, L6 
\bibitem[Athanassoula et al.(2010)]{atha10} Athanassoula, E., Romero-G{\'o}mez, M., Bosma, A., \& Masdemont, J.~J.\ 2010, \mnras, 407, 1433 
\bibitem[Block et al.(1994)]{blo94} Block, D.~L., Bertin, G., Stockton, A., et al.\ 1994, \aap, 288, 365 
\bibitem[Bruzual \& Charlot(2003)]{bc03} Bruzual, G., \& Charlot, S.\ 2003, \mnras, 344, 1000 
\bibitem[Clarke \& Gittins(2006)]{cla06} Clarke, C., \& Gittins, D.\ 2006, \mnras, 371, 530 
\bibitem[Charlot \& Fall(2000)]{cha00} Charlot, S., \& Fall, S.~M.\ 2000, \apj, 539, 718 
\bibitem[Considere \& Athanassoula(1982)]{con82} Considere, S., \& Athanassoula, E.\ 1982, \aap, 111, 28
\bibitem[Considere \& Athanassoula(1988)]{con88} Considere, S., \& Athanassoula, E.\ 1988, \aaps, 76, 365 
\bibitem[da Cunha et al.(2008)]{daC08} da Cunha, E., Charlot, S., \& Elbaz, D.\ 2008, \mnras, 388, 1595
\bibitem[Davis et al.(2012)]{dav12} Davis, B.~L., Berrier, J.~C., Shields, D.~W., et al.\ 2012, \apjs, 199, 33 
\bibitem[de Vaucouleurs et al.(1991)]{dev91} de Vaucouleurs, G., de Vaucouleurs, A., Corwin, H.~G., Jr., et al.\ 1991, Third Reference Catalogue of Bright Galaxies. (Berlin: Springer), (RC3)
\bibitem[Dobbs \& Bonnell(2008)]{dobbo08} Dobbs, C.~L., \& Bonnell, I.~A.\ 2008, \mnras, 385, 1893 
\bibitem[Dobbs \& Price(2008)]{dobpr08} Dobbs, C.~L., \& Price, D.~J.\ 2008, \mnras, 383, 497 
\bibitem[Donner \& Thomasson(1994)]{don94} Donner, K.~J., \& Thomasson, M.\ 1994, \aap, 290, 785 
\bibitem[Efremov(1985)]{efr85} Efremov, Y.~N.\ 1985, Soviet Astronomy Letters, 11, 69 
\bibitem[Efremov(2010)]{efr10} Efremov, Y.~N.\ 2010, \mnras, 405, 1531 
\bibitem[Efremov(2011)]{efr11} Efremov, Y.~N.\ 2011, Astronomy Reports, 55, 108 
\bibitem[Egusa et al.(2009)]{egu09} Egusa, F., Kohno, K., Sofue, Y., Nakanishi, H., \& Komugi, S.\ 2009, \apj, 697, 1870 
\bibitem[Ferreras et al.(2012)]{fer12} Ferreras, I., Cropper, M., Kawata, D., Page, M., \& Hoversten, E.~A.\ 2012, \mnras, 424, 1636 
\bibitem[Foyle et al.(2011)]{foy11} Foyle, K., Rix, H.-W., Dobbs, C.~L., Leroy, A.~K., \& Walter, F.\ 2011, \apj, 735, 101 
\bibitem[Fujii et al.(2011)]{fuj11} Fujii, M.~S., Baba, J., Saitoh, T.~R., et al.\ 2011, \apj, 730, 109 
\bibitem[Gerola \& Seiden(1978)]{ger78} Gerola, H., \& Seiden, P.~E.\ 1978, \apj, 223, 129 
\bibitem[Gittins \& Clarke(2004)]{git04} Gittins, D.~M., \& Clarke, C.~J.\ 2004, \mnras, 349, 909 
\bibitem[Goodwin \& Bastian(2006)]{goo06} Goodwin, S.~P., \& Bastian, N.\ 2006, \mnras, 373, 752
\bibitem[Gonz\'alez \& Graham(1996)]{gon96} Gonz\'alez, R. A., \& Graham, J. R. 1996, \apj , 460, 651 (GG96)
\bibitem[Gonz{\'a}lez-L{\'o}pezlira et al.(2012)]{gon12} Gonz{\'a}lez-L{\'o}pezlira, R.~A., Pflamm-Altenburg, J., \& Kroupa, P.\ 2012, \apj, 761, 124 
\bibitem[Grand et al.(2012a)]{gra12a} Grand, R.~J.~J., Kawata, D., \& Cropper, M.\ 2012a, \mnras, 421, 1529 
\bibitem[Grand et al.(2012b)]{gra12b} Grand, R.~J.~J., Kawata, D., \& Cropper, M.\ 2012b, \mnras, 426, 167 
\bibitem[Grosb{\o}l \& Patsis(1998)]{gro98} Grosb{\o}l, P.~J., \& Patsis, P.~A.\ 1998, \aap, 336, 840 
\bibitem[Grosb{\o}l \& Dottori(2009)]{gro09} Grosb{\o}l, P., \& Dottori, H.\ 2009, \aap, 499, L21 
\bibitem[Grosb{\o}l et al.(2006)]{gros06} Grosb{\o}l, P., Dottori, H., \& Gredel, R.\ 2006, \aap, 453, L25 
\bibitem[Hozumi(2003)]{hoz03} Hozumi, S.\ 2003, in Galaxies and Chaos, ed. G. Contopoulos \& N. Voglis (Lecture Notes in Physics, Vol. 626; Berlin: Springer), 380
\bibitem[James \& Seigar(1999)]{jam99} James, P.~A., \& Seigar, M.~S.\ 1999, \aap, 350, 791 
\bibitem[Lada \& Lada(2003)]{lad03} Lada, C.~J., \& Lada, E.~A.\ 2003, \araa, 41, 57
\bibitem[Lin \& Shu(1964)]{lin64} Lin, C.~C., \& Shu, F.~H.\ 1964, \apj, 140, 646
\bibitem[Lindblad(1963)]{lind63} Lindblad, B.\ 1963, Stockholms Observatoriums Annaler, 22, 5
\bibitem[Mart{\'{\i}}nez-Garc{\'{\i}}a et al.(2009a)]{mar09a} Mart{\'{\i}}nez-Garc{\'{\i}}a, E.~E., Gonz{\'a}lez-L{\'o}pezlira, R.~A., \& Bruzual-A, G.\ 2009a, \apj, 694, 512 
\bibitem[Mart{\'{\i}}nez-Garc{\'{\i}}a et al.(2009b)]{mar09b} Mart{\'{\i}}nez-Garc{\'{\i}}a, E.~E., Gonz{\'a}lez-L{\'o}pezlira, R.~A., \& G{\'o}mez, G.~C.\ 2009b, \apj, 707, 1650 
\bibitem[Mart{\'{\i}}nez-Garc{\'{\i}}a \& Gonz{\'a}lez-L{\'o}pezlira(2011)]{mar11} Mart{\'{\i}}nez-Garc{\'{\i}}a, E.~E., \& Gonz{\'a}lez-L{\'o}pezlira, R.~A.\ 2011, \apj, 734, 122 
\bibitem[Mart{\'{\i}}nez-Garc{\'{\i}}a(2012)]{mar12} Mart{\'{\i}}nez-Garc{\'{\i}}a, E.~E.\ 2012, \apj, 744, 92 
\bibitem[Meidt et al.(2009)]{mei09} Meidt, S.~E., Rand, R.~J., \& Merrifield, M.~R.\ 2009, \apj, 702, 277 
\bibitem[Merrifield et al.(2006)]{mer06} Merrifield, M.~R., Rand, R.~J., \& Meidt, S.~E.\ 2006, \mnras, 366, L17 
\bibitem[Mould et al.(2000)]{mou00} Mould, J. R.~et al.\ 2000, \apj, 529, 786
\bibitem[Mueller \& Arnett(1976)]{mu76} Mueller, M.~W., \& Arnett, W.~D.\ 1976, \apj, 210, 670 
\bibitem[Patsis et al.(2001)]{pat01} Patsis, P.~A., H{\'e}raudeau, P., \& Grosb{\o}l, P.\ 2001, \aap, 370, 875 
\bibitem[Paturel et al.(2000)]{pat00} Paturel, G., Fang, Y., Petit, C., Garnier, R., \& Rousseau, J.\ 2000, \aaps, 146, 19 
\bibitem[Paturel et al.(2003)]{pat03} Paturel, G., Theureau, G., Bottinelli, L., Gouguenheim, L., Coudreau-Durand, N., Hallet, N., Petit, C. 2003, \aap, 412, 57
\bibitem[Pflamm-Altenburg \& Kroupa(2008)]{pfl08} Pflamm-Altenburg, J., \& Kroupa, P.\ 2008, \nat, 455, 641 
\bibitem[Pflamm-Altenburg et al.(2009)]{pfla09} Pflamm-Altenburg, J., Weidner, C., \& Kroupa, P.\ 2009, \mnras, 395, 394
\bibitem[Pilbratt et al.(2010)]{pil10} Pilbratt, G.~L., Riedinger, J.~R., Passvogel, T., et al.\ 2010, \aap, 518, L1 
\bibitem[Puerari \& Dottori(1992)]{pue92} Puerari, I., \& Dottori, H.~A.\ 1992, \aaps, 93, 469 
\bibitem[Puerari \& Dottori(1997)]{pue97} Puerari, I., \& Dottori, H.\ 1997, \apjl, 476, L73 
\bibitem[Rix \& Rieke(1993)]{rix93} Rix, H.W., \& Rieke, M. J. 1993, \apj, 418, 123
\bibitem[Rhoads(1998)]{rho98} Rhoads, J.~E.\ 1998, \aj, 115, 472 
\bibitem[Roberts(1969)]{rob69} Roberts, W.~W.\ 1969, \apj, 158, 123 
\bibitem[Roca-F{\`a}brega et al.(2013)]{roc13} Roca-F{\`a}brega, S., Valenzuela, O., Figueras, F., et al. 2013, \mnras, submitted
\bibitem[Ringermacher \& Mead(2009)]{rin09} Ringermacher, H.~I., \& Mead, L.~R.\ 2009, \mnras, 397, 164 
\bibitem[S{\'a}nchez-Gil et al.(2011)]{san11} S{\'a}nchez-Gil, M.~C., Jones, D.~H., P{\'e}rez, E., et al.\ 2011, \mnras, 415, 753 
\bibitem[Saraiva Schroeder et al.(1994)]{sar94} Saraiva Schroeder, M.~F., Pastoriza, M.~G., Kepler, S.~O., \& Puerari, I.\ 1994, \aaps, 108, 41 
\bibitem[Savchenko(2012)]{sav12} Savchenko, S.~S.\ 2012, Astrophysical Bulletin, 67, 310 
\bibitem[Scarano et al.(2011)]{sca11} Scarano, S., Jr., L{\'e}pine, J.~R.~D., \& Marcon-Uchida, M.~M.\ 2011, \mnras, 412, 1741 
\bibitem[Scarano \& L{\'e}pine(2013)]{sca13} Scarano, S., \& L{\'e}pine, J.~R.~D.\ 2013, \mnras, 428, 625 
\bibitem[Schlegel et al.(1998)]{sch98} Schlegel, D.~J., Finkbeiner, D.~P., \& Davis, M.\ 1998, \apj, 500, 525 
\bibitem[Sellwood(2011)]{sel11} Sellwood, J.~A.\ 2011, \mnras, 410, 1637 
\bibitem[Speights \& Westpfahl(2012)]{spe12} Speights, J.~C., \& Westpfahl, D.~J.\ 2012, \apj, 752, 52 
\bibitem[Skrutskie et al.(2006)]{skr06} Skrutskie, M.~F., Cutri, R.~M., Stiening, R., et al.\ 2006, \aj, 131, 1163 
\bibitem[Tamburro et al.(2008)]{tam08} Tamburro, D., Rix, H.-W., Walter, F., et al.\ 2008, \aj, 136, 2872 
\bibitem[Thuan \& Gunn(1976)]{thu76} Thuan, T. X., Gunn, J. E. 1976, \pasp, 88, 543
\bibitem[Wada et al.(2011)]{wada11} Wada, K., Baba, J., \& Saitoh, T.~R.\ 2011, \apj, 735, 1 
\bibitem[Wade et al.(1979)]{wade79} Wade, R. A., Hoessel, J. G., Elias, J. H., Huchra, J. P. 1979, \pasp, 91, 35
\bibitem[Wielen(1977)]{wie77} Wielen, R.\ 1977, \aap, 60, 263 
\bibitem[Zhang(1998)]{zha98} Zhang, X.\ 1998, \apj, 499, 93 
\bibitem[Zibetti et al.(2009)]{zib09} Zibetti, S., Charlot, S., \& Rix, H.-W.\ 2009, \mnras, 400, 1181
\bibitem[Zibetti(2009)]{zib09b} Zibetti, S.\ 2009, arXiv:0911.4956 

\end{thebibliography}
\end{document}